\documentclass[aps,prd,preprintnumbers,showpacs,superscriptaddress,tighten,nofootinbib,preprint]{revtex4}
\usepackage{graphicx}
\usepackage{amsmath}

\begin{document}

\title{Effects of nonstandard neutrino interactions at PINGU}

\author{Tommy Ohlsson}

\email{tohlsson@kth.se}

\affiliation{Department of Theoretical Physics, School of
Engineering Sciences, KTH Royal Institute of Technology, AlbaNova
University Center, 106 91 Stockholm, Sweden}

\author{He Zhang}

\email{he.zhang@mpi-hd.mpg.de}

\affiliation{Max-Planck-Institut f\"{u}r Kernphysik, Saupfercheckweg
1, 69117 Heidelberg, Germany}

\author{Shun Zhou}
\email{shunzhou@kth.se}

\affiliation{Department of Theoretical Physics, School of
Engineering Sciences, KTH Royal Institute of Technology, AlbaNova
University Center, 106 91 Stockholm, Sweden}

\begin{abstract}
Neutrino oscillation experiments in the past decades have greatly
improved our knowledge on neutrinos by measuring the fundamental
neutrino parameters. The ongoing and upcoming neutrino oscillation
experiments are intended to pin down the neutrino mass hierarchy and
to discover the leptonic CP violation. By means of neutrino
oscillograms, we analyze the impact of non-standard neutrino
interactions on neutrino oscillations in the Earth matter. The
standard neutrino oscillation probabilities may be significantly
changed by non-standard interaction parameters, and in particular,
the CP-violating effects in the energy range $E= 1 \sim 20~{\rm
GeV}$ are greatly enhanced. In addition, the event rates of muon
neutrinos in the proposed huge atmospheric neutrino experiment,
PINGU at the South Pole, have been estimated in the presence of
non-standard neutrino interactions. It has been found that the PINGU
experiment has very good sensitivities to the non-standard neutrino
interaction parameters.
\end{abstract}


\maketitle

\section{Introduction}

Today, the phenomenon of neutrino oscillations is considered to be
the standard and leading order mechanism for neutrino flavor
transitions, providing strong evidence that neutrinos are massive
and lepton flavors are mixed, which leads to physics beyond the
Standard Model (SM) of particle physics \cite{Beringer:1900zz}.
However, although it is widely accepted in the particle physics
community that this phenomenon stems from a non-trivial structure of
leptonic flavor mixing, the so-called non-standard neutrino
interactions (NSIs), which are considered to be sub-leading order
effects to standard neutrino oscillations, may still affect neutrino
flavor transitions in a significant way \cite{Ohlsson:2012kf}. The
concept of NSIs is presently the most studied description for
effects beyond the standard paradigm of neutrino oscillations. In
fact, dimension-six and higher-order operators exist in various
theoretical extensions of the SM, which include e.g. seesaw models,
$R$-parity violating supersymmetric models, left-right symmetric
models, grand unification theories, and extra dimensions. Basically,
all modern extensions could give rise to NSIs. Therefore, the
investigation of NSIs could reveal additional new physics behind
neutrino flavor transitions. In addition, it plays an important
complementary role to direct searches of physics beyond the SM at
colliders such as the LHC.

For example, NSI effects have previously been studied for the
accelerator-based neutrino oscillation experiments MINOS and OPERA
\cite{Blennow:2007pu,Blennow:2008eb}, atmospheric neutrino
experiments \cite{Fornengo:2001pm, Huber:2001zw, Mitsuka:2011ty},
reactor neutrino experiments \cite{Kopp:2007ne,
Ohlsson:2008gx,Leitner}, and a future neutrino factory
\cite{Kopp:2007mi, Ribeiro:2007ud, Kopp:2008ds, Meloni:2009cg,
Coloma:2011rq}. In this work, we investigate neutrino flavor
transition probabilities based on standard neutrino oscillations and
NSIs as sub-leading effects. Note that we only consider so-called
propagation (or matter) NSIs, which are parameterized by different
NSI parameters, and not source and detector NSIs. Especially, we
derive oscillation probabilities for the $\nu_e \to \nu_\mu$ and
$\nu_\mu \to \nu_\mu$ channels that are important for atmospheric
neutrino oscillations, and study the impact of various NSI
parameters on these probabilities, in both cases of the normal
neutrino mass hierarchy (NH) and inverted neutrino mass hierarchy
(IH). In addition, we estimate the number of atmospheric neutrino
events in the future PINGU experiment at the South Pole, which has
been recently shown to have great potential for determining the
neutrino mass hierarchy \cite{Akhmedov:2012ah, Agarwalla:2012uj,
Franco:2013in, Ribordy:2013xea}. Very good sensitivities to the NSI
parameters are expected at this experiment.

The remaining part of our paper is organized as follows. In Sec. II,
we will review the formalism of three-flavor neutrino oscillations
in the presence of standard and non-standard matter effects, and
present the parameter mappings between the leptonic mixing matrix in
vacuum and that in matter. The analytical approximate formulas of
oscillation probabilities are derived for the appearance channel
$\nu_e \to \nu_\mu$ and disappearance channel $\nu_\mu \to \nu_\mu$.
Sec. III is devoted to numerical analyses of the NSI effects on the
neutrino oscillation probabilities with and without intrinsic CP
violation. Using neutrino oscillograms, we further point out the
important regions in the plane of zenith angles and neutrino
energies, where the NSI effects are most significant. Taking into
account the NSIs, we estimate the event rate of atmospheric muon
neutrinos at the PINGU detector in Sec. IV. It turns out that the
PINGU has very good sensitivities to the NSI parameters. Finally, we
conclude in Sec. V.

\section{Neutrino Oscillations with NSIs}

In general, NSIs can be present for neutrino production,
propagation, and detection. We will concentrate on the non-standard
matter effects for neutrino propagation, which should be relevant
for atmospheric and long-baseline neutrino oscillation experiments
in particular. In this section, we recall the general formulation of
three-flavor neutrino oscillations in the presence of standard and
non-standard matter effects. In this case, neutrino flavor
transitions are determined by the effective Hamiltonian
\begin{equation}
{\cal H}(x) = {\cal H}_0 + {\cal H}_{\rm m}(x) + {\cal H}_{\rm
NSI}(x) \; ,
\end{equation}
where the vacuum Hamiltonian is given by
\begin{equation}
{\cal H}_0 = \frac{1}{2E} U \left(\begin{matrix}0 & 0 & 0 \cr 0 &
\Delta m^2_{21} & 0 \cr 0 & 0 & \Delta m^2_{31}
\end{matrix}\right) U^\dagger \; ,
\end{equation}
the standard matter potential is
\begin{equation}
{\cal H}_{\rm m}(x) = V_{\rm CC} \left(\begin{matrix}1 & 0 & 0 \cr 0
& 0 & 0 \cr 0 & 0 & 0\end{matrix}\right) \; ,
\end{equation}
with $V_{\rm CC}=\sqrt{2} G_{\rm F} N_e(x)$, and the non-standard
matter potential is parametrized by
\begin{equation}
{\cal H}_{\rm NSI}(x) = V_{\rm CC}
\left(\begin{matrix}\varepsilon_{ee} & \varepsilon_{e\mu} &
\varepsilon_{e\tau} \cr \varepsilon^*_{e\mu} & \varepsilon_{\mu \mu}
& \varepsilon_{\mu \tau} \cr \varepsilon^*_{e\tau} &
\varepsilon^*_{\mu \tau} & \varepsilon_{\tau
\tau}\end{matrix}\right) \; ,
\end{equation}
where $\varepsilon^{}_{\alpha \beta}$ are real (for $\alpha =
\beta$) or complex (for $\alpha \neq \beta$) constants, i.e., the
so-called NSI parameters. Note that $G^{}_{\rm F}$ is the Fermi
constant and $N^{}_e(x)$ is the electron number density in matter.

The exact oscillation probability is given by $P_{\alpha\beta} =
|S_{\beta\alpha}(x,x_0)|^2$, where $S_{\beta\alpha}(x,x_0)$ is the
evolution matrix satisfying the Schr{\"o}dinger-like equation,
\begin{eqnarray}
{\rm i}\frac{\rm d}{{\rm d}x}|\nu(x)\rangle = {\cal H}(x) |
\nu(x)\rangle \; ,
\end{eqnarray}
and can be obtained by solving the above equation as
\begin{eqnarray}
S(x,x_0) = \exp\left[-{\rm i} \int^{x}_{x_0} {\cal H}(x^\prime)
~{\rm d} x^\prime\right] \; .
\end{eqnarray}
Since the standard matter potential ${\cal H}_{\rm m}$ is invariant
under any rotation in the $2$-$3$ plane, it is sometimes convenient
to work in a new flavor basis $(\nu^{}_e, \tilde{\nu}^{}_2,
\tilde{\nu}^{}_3)^T = U^{}_{23} (\nu^{}_e, \nu^{}_\mu,
\nu^{}_\tau)^T$ with $U^{}_{23} \equiv O^{}_{23} I^{}_\delta$. Note
that the leptonic mixing matrix can be parametrized as $U =
O^{}_{23} I^{}_\delta O^{}_{13} I^\dagger_{\delta} O^{}_{12}$, where
$O^{}_{ij}$ denotes the rotation in the $i$-$j$ plane with a
rotation angle $\theta^{}_{ij}$ and $I^{}_\delta \equiv {\rm
diag}(1, 1, e^{{\rm i}\delta})$ with $\delta$ being the leptonic
Dirac CP-violating phase. If neutrinos are Majorana particles, there
will be two additional CP-violating phases in $U$. However, these
two leptonic Majorana CP-violating phases are irrelevant for
neutrino oscillations both in vacuum and in matter.

Now, in this basis, the standard Hamiltonian can be written as
\begin{equation}
\tilde{\cal H}_{\rm SD}(x) = \frac{1}{2E} O^{}_{13} O^{}_{12}
\left(\begin{matrix}0 & 0 & 0 \cr 0 & \Delta m^2_{21} & 0 \cr 0 & 0
& \Delta m^2_{31} \end{matrix}\right) O^T_{12} O^T_{13} + {\cal
H}_{\rm m}(x) \; ,
\end{equation}
where $\tilde{\cal H}_{\rm SD}(x) = U^\dagger_{23} {\cal H}_{\rm
SD}(x) U_{23}$ with ${\cal H}_{\rm SD}(x) = {\cal H}_0 + {\cal
H}_{\rm m}(x)$, while the non-standard matter potential turns out to
be
\begin{equation}
\tilde{\cal H}^{}_{\rm NSI}(x) = V^{}_{\rm CC}
\left(\begin{matrix}\varepsilon^{}_{ee} &
\tilde{\varepsilon}^{}_{e\mu} & \tilde{\varepsilon}^{}_{e\tau} \cr
\tilde{\varepsilon}^*_{e\mu} & \tilde{\varepsilon}^{}_{\mu \mu} &
\tilde{\varepsilon}^{}_{\mu \tau} \cr \tilde{\varepsilon}^*_{e\tau}
& \tilde{\varepsilon}^*_{\mu \tau} & \tilde{\varepsilon}^{}_{\tau
\tau}\end{matrix}\right) \; ,
\end{equation}
with the modified NSI parameters
\begin{eqnarray}
\tilde{\varepsilon}^{}_{e\mu} &=& \varepsilon^{}_{e\mu} c^{}_{23} -
\varepsilon^{}_{e\tau} s^{}_{23} \; , \nonumber \\
\tilde{\varepsilon}^{}_{e\tau} &=& (\varepsilon^{}_{e\mu} s^{}_{23}
+
\varepsilon^{}_{e\tau} c^{}_{23}) e^{{\rm i}\delta} \; , \nonumber \\
\tilde{\varepsilon}^{}_{\mu \mu} &=& (\varepsilon^{}_{\mu \mu}
c^2_{23} + \varepsilon^{}_{\tau \tau} s^2_{23}) - 2s^{}_{23}
c^{}_{23} {\rm
Re}[\varepsilon^{}_{\mu \tau}] \; , \nonumber \\
\tilde{\varepsilon}^{}_{\tau \tau} &=& (\varepsilon^{}_{\mu \mu}
s^2_{23} + \varepsilon^{}_{\tau \tau} c^2_{23}) + 2s^{}_{23}
c^{}_{23} {\rm
Re}[\varepsilon^{}_{\mu \tau}] \; , \nonumber \\
\tilde{\varepsilon}^{}_{\mu \tau} &=& \left[(\varepsilon^{}_{\mu
\tau} c^2_{23} - \varepsilon^*_{\mu \tau} s^2_{23}) +
(\varepsilon^{}_{\mu \mu} - \varepsilon^{}_{\tau \tau}) s^{}_{23}
c^{}_{23}\right] e^{{\rm i} \delta} \; .
\end{eqnarray}
Hence, in this basis, the effective Hamiltonian is
\begin{equation}
\tilde{\cal H}(x) = \Delta_{31} \left[\left(\begin{matrix} s^2_{12}
c^2_{13} \alpha + s^2_{13} & s^{}_{12} c^{}_{12} c^{}_{13} \alpha &
s^{}_{13} c^{}_{13} (1-s^2_{12}\alpha) \cr s^{}_{12} c^{}_{12}
c^{}_{13} \alpha & c^2_{12} \alpha & - s^{}_{12} c^{}_{12} s^{}_{13}
\alpha \cr s^{}_{13} c^{}_{13} (1-s^2_{12}\alpha) & - s^{}_{12}
c^{}_{12} s^{}_{13} \alpha & s^2_{12} s^2_{13} \alpha + c^2_{13}
\end{matrix}\right) + A \left( \begin{matrix} 1+ \varepsilon^{}_{ee} &
\tilde{\varepsilon}^{}_{e\mu} & \tilde{\varepsilon}^{}_{e\tau} \cr
\tilde{\varepsilon}^*_{e\mu} & \tilde{\varepsilon}^{}_{\mu \mu} &
\tilde{\varepsilon}^{}_{\mu \tau} \cr \tilde{\varepsilon}^*_{e\tau}
& \tilde{\varepsilon}^*_{\mu \tau} & \tilde{\varepsilon}^{}_{\tau
\tau}\end{matrix}\right)\right] \; , \nonumber
\end{equation}
where $\Delta_{31} \equiv \Delta m^2_{31}/(2E)$, $\alpha \equiv
\Delta m^2_{21}/\Delta m^2_{31}$ and $A \equiv V_{\rm
CC}/\Delta_{31}$. Additionally, note that $s^{}_{ij} \equiv \sin
\theta^{}_{ij}$ and $\cos \theta^{}_{ij} \equiv c^{}_{ij}$ have been
defined. The evolution matrix $\tilde{S}(x,x^{}_0)$ in this basis is
related to $S(x,x_0)$ in the flavor basis via the unitary
transformation $S(x,x^{}_0) = U^{}_{23} \tilde{S}(x,x_0)
U^\dagger_{23}$. The oscillation probabilities of antineutrinos can
be obtained through the replacements $A \to - A$ and $\delta \to
-\delta$.

\subsection{Parameter Mappings}

Now, following Refs. \cite{Akhmedov:2001kd} and
\cite{Akhmedov:2004ny}, we use a perturbation method to derive the
effective neutrino masses $\tilde{m}^2_{i}$ and leptonic mixing
matrix $U^{\rm m}$ in matter. Note that current neutrino oscillation
data indicate $\alpha \sim \sqrt{2} s^2_{13} \approx 0.03$.
Therefore, we keep the terms of the same order $\alpha$ and
$s^2_{13}$, and ignore all other higher-order contributions, such as
$\alpha s^{}_{13}$, $\alpha^2$, and $\alpha s^2_{13}$. Thus, we
write $\tilde{\cal H} = {\cal M} \cdot \Delta^{}_{31}$ and introduce
${\cal M} \equiv {\cal M}^{(0)} + {\cal M}^{(1)}$:
\begin{eqnarray}
{\cal M}^{(0)} &=& \left(\begin{matrix} s^2_{13} + A & 0 & s^{}_{13}
c^{}_{13} \cr 0 & 0 & 0 \cr s^{}_{13} c^{}_{13} & 0 & c^2_{13}
\end{matrix}\right) \; , \\
{\cal M}^{(1)} &=& \left(\begin{matrix} s^2_{12} \alpha + A
\varepsilon^{}_{ee} & s^{}_{12} c^{}_{12} \alpha + A
\tilde{\varepsilon}^{}_{e\mu} & A \tilde{\varepsilon}^{}_{e\tau} \cr
s^{}_{12} c^{}_{12} \alpha + A \tilde{\varepsilon}^*_{e\mu} &
c^2_{12} \alpha + A \tilde{\varepsilon}^{}_{\mu\mu} & A
\tilde{\varepsilon}^{}_{\mu\tau} \cr A \tilde{\varepsilon}^*_{e\tau}
& A \tilde{\varepsilon}^*_{\mu\tau} & A
\tilde{\varepsilon}^{}_{\tau\tau}
\end{matrix}\right) \; ,
\end{eqnarray}
where ${\cal M}^{(0)}$ corresponds exactly to the standard
Hamiltonian $\tilde{\cal H}^{}_{\rm SD}$ in the two-flavor limit
with $\alpha = 0$, while ${\cal M}^{(1)}$ incorporates the
corrections from $\alpha$ and the NSI parameters. Obviously, ${\cal
M}^{(0)}$ can be diagonalized by a rotation in the $1$-$3$ plane,
i.e.,
\begin{equation}
U^{(0)} = \left(\begin{matrix} \cos \hat{\theta}_{13} & 0 & \sin
\hat{\theta}^{}_{13} \cr 0 & 1 & 0 \cr -\sin \hat{\theta}^{}_{13} &
0 & \cos \hat{\theta}^{}_{13}
\end{matrix}\right) \;
\end{equation}
with the effective mixing angle $\hat{\theta}_{13}$ given by
\begin{equation}
\tan 2\hat{\theta}^{}_{13} = \frac{\sin 2\theta^{}_{13}}{\cos
2\theta^{}_{13} - A} \; .
\end{equation}
In the following, we choose $\hat{\theta}^{}_{13}$ to be defined in
the first quadrant \footnote{Note that it is also possible to define
$\hat{\theta}_{13}$ to be in $[0, \pi/4]$ by properly arranging the
eigenvalues and the corresponding eigenvectors, as shown in
Ref.~\cite{Freund:2001pn}.}. Namely, $\hat{\theta}^{}_{13} \in [0,
\pi/4]$ for $A < \cos 2\theta^{}_{13}$ and $\hat{\theta}^{}_{13} \in
[\pi/4, \pi/2]$ for $A > \cos 2\theta^{}_{13}$.  Therefore, we
obtain \cite{Akhmedov:2004ny}
\begin{eqnarray}
\sin^2\hat{\theta}^{}_{13} = \frac{\hat{C} - \cos 2\theta^{}_{13} +
A}{2\hat{C}} \; , ~~~ \cos^2\hat{\theta}^{}_{13} = \frac{\hat{C} +
\cos 2\theta^{}_{13} - A}{2\hat{C}} \; ,
\end{eqnarray}
with $\hat{C} \equiv \sqrt{(\cos 2\theta^{}_{13} - A)^2 + \sin^2
2\theta^{}_{13}}$. Two other useful relations can readily be derived
from Eq.~(14)
\begin{equation}
\sin 2\hat{\theta}^{}_{13} = \frac{\sin 2\theta^{}_{13}}{\hat{C}} \;
, ~~~~ \cos 2\hat{\theta}^{}_{13} = \frac{\cos 2 \theta^{}_{13} -
A}{\hat{C}} \; ,
\end{equation}
implying $\sin \hat{\theta}_{13} = \sin \theta_{13}/(1-A)$ at the
leading order of $\sin \theta_{13}$. In the limit of $\alpha = 0$
and in the absence of NSIs, $\hat{\theta}_{13}$ is just the
effective mixing angle in matter.

Furthermore, the eigenvalues to zeroth order are given by
\begin{eqnarray}
\lambda^{(0)}_1 &=& \frac{1}{2}\left(1+A - \hat{C}\right) \; ,
\nonumber \\
\lambda^{(0)}_2 &=& 0 \; , \nonumber \\
\lambda^{(0)}_3 &=& \frac{1}{2}\left(1+A + \hat{C}\right) \; ,
\end{eqnarray}
and the eigenvalues to first order are $\lambda^{(1)}_i =
\tilde{\cal M}^{(1)}_{ii}$ with $\tilde{\cal M}^{(1)} =
{U^{(0)}}^\dagger {\cal M}^{(1)} U^{(0)}$. It is straightforward to
show
\begin{eqnarray}
\lambda^{(1)}_1 &=& \alpha \cos^2\hat{\theta}_{13} s^2_{12} + A
\left[\cos^2\hat{\theta}_{13} \varepsilon^{}_{ee} - \sin
2\hat{\theta}^{}_{13} {\rm Re}(\tilde{\varepsilon}^{}_{e\tau}) +
\sin^2\hat{\theta}_{13} \tilde{\varepsilon}^{}_{\tau\tau}\right] \;
,
\nonumber \\
\lambda^{(1)}_2 &=& \alpha c^2_{12}  + A
\tilde{\varepsilon}^{}_{\mu\mu} \; ,  \nonumber \\
\lambda^{(1)}_3 &=& \alpha \sin^2\hat{\theta}_{13} s^2_{12} + A
\left[\sin^2\hat{\theta}_{13} \varepsilon^{}_{ee} + \sin
2\hat{\theta}^{}_{13} {\rm Re}(\tilde{\varepsilon}^{}_{e\tau}) +
\cos^2\hat{\theta}_{13} \tilde{\varepsilon}^{}_{\tau\tau}\right] \;.
\end{eqnarray}
The corrections to the eigenvectors are given by
\begin{eqnarray}
U^{(1)}_i = \sum_{j\neq i} \frac{\tilde{\cal
M}^{(1)}_{ji}}{\lambda^{(0)}_i - \lambda^{(0)}_j} U^{(0)}_j \; ,
\end{eqnarray}
where $U^{(1)}_i$ and $U^{(0)}_j$ stand for the column vectors of
the matrices $U^{(1)}$ and $U^{(0)}$, respectively. The effective
neutrino masses in matter are determined by $\tilde{m}^2_i = m^2_1 +
\Delta m^2_{31} [\lambda^{(0)}_i + \lambda^{(1)}_i]$, while the
effective leptonic mixing matrix is $U^{\rm m} = U^{}_{23}
\left[U^{(0)} + U^{(1)}\right]$. In the absence of NSIs, there are
two resonances, i.e., $A = \alpha$ and $A = \cos 2 \theta^{}_{13}$.
For neutrino energies $E > 1~{\rm GeV}$ and matter densities $\rho =
3~{\rm g}/{\rm cm}^3$ in the Earth crust, we have $A > \alpha$, so
only the resonance $A = \cos 2\theta^{}_{13}$ is relevant
\cite{Freund:2001pn}. Now, after some lengthy computations, we find
that
\begin{eqnarray}
U^{\rm m}_{e3} &=& \sin \hat{\theta}^{}_{13} + \frac{\cos
\hat{\theta}^{}_{13}}{2\hat{C}} \left\{\sin 2\hat{\theta}^{}_{13}
\left[\alpha s^2_{12} + A(\varepsilon^{}_{ee} -
\tilde{\varepsilon}^{}_{\tau\tau})\right] + 2A\left[\cos
2\hat{\theta}^{}_{13} {\rm Re}(\tilde{\varepsilon}^{}_{e\tau}) +
{\rm i} {\rm Im}(\tilde{\varepsilon}^{}_{e\tau})\right]\right\} \; ,
\nonumber \\
U^{\rm m}_{e2} &=& -\frac{c^{}_{13}}{2A} \alpha \sin 2\theta^{}_{12}
- \tilde{\varepsilon}^{}_{e\mu} + \frac{\tan \theta^{}_{13}}{A}
\tilde{\varepsilon}^*_{\mu \tau} \; , \nonumber \\
U^{\rm m}_{\mu 3} &=& s^{}_{23} \cos \hat{\theta}^{}_{13} e^{{\rm
i}\delta} \left\{1 - \frac{A \tan \hat{\theta}^{}_{13}}{\hat{C}}
\left[\cos 2\hat{\theta}^{}_{13} {\rm
Re}(\tilde{\varepsilon}^{}_{e\tau}) + {\rm i} {\rm
Im}(\tilde{\varepsilon}^{}_{e\tau})\right] \right\} + \frac{\alpha
c^{}_{23} \sin 2\theta^{}_{12} \sin
\hat{\theta}^{}_{13}}{1+A+\hat{C}} \; , \nonumber \\
\end{eqnarray}
which reproduce the well-known results in the limit of vanishing NSI
parameters \cite{Freund:2001pn,Meloni:2009ia}. Our results in
Eq.~(19) differ from those in Ref.~\cite{Meloni:2009ia} by including
higher-order corrections from $\alpha$ and $s^{}_{13}$. The
parameter mapping is not valid in the resonance region, where the
perturbation theory breaks down. Furthermore, the mixing matrix
elements in matter could be divergent, so we will calculate the
oscillation probabilities that should be well-defined in general,
and particularly in the resonance region. As pointed out in
Ref.~\cite{Freund:2001pn}, the mixing angle $\theta_{12}$ can be
arbitrary in the limit of $\alpha = 0$ and in the absence of the
NSIs, so the mapping for $U^{\rm m}_{e2}$ in Eq.~(19) cannot be
taken seriously. For neutrino energies $E > 1~{\rm GeV}$, only the
effective mixing angle $\tilde{\theta}_{13}$, which is determined by
$U^{\rm m}_{e3}$ in the standard parametrization
\cite{Beringer:1900zz}, is crucially important for neutrino
oscillations in the Earth matter. Note that the true effective
mixing angle $\tilde{\theta}_{13}$ differs from $\hat{\theta}_{13}$
in the contributions from $\alpha$ and the NSI parameters.

\subsection{Oscillation Probabilities}

In practice, the oscillation probabilities can be computed using
perturbation theory based on small quantities, e.g., the smallest
mixing angle $\theta_{13}$, the ratio of the two mass-squared
differences $\alpha$, and the NSI parameters $\varepsilon_{\alpha
\beta}$. Following Ref.~\cite{Akhmedov:2004ny}, one can explicitly
decompose the effective Hamiltonian as ${\cal H}(x) = {\cal
H}^\prime_0(x) + {\cal H}_{\rm I}(x)$, where ${\cal H}^\prime_0(x)$
is the zeroth order in the small parameters and ${\cal H}_{\rm
I}(x)$ includes the higher-order contributions. Then, to first
order, the evolution matrix is approximately given by
\begin{eqnarray}
S(x,x_0) \simeq S_0(x,x_0) - {\rm i} S_0(x,x_0) \int^x_{x_0} \left[
S_0(x^\prime,x_0)^{-1} {\cal H}_{\rm I}(x^\prime) S_0(x^\prime,x_0)
\right] {\rm d} x^\prime \; ,
\end{eqnarray}
where the zeroth-order evolution matrix $S_0(x,x_0)$ is determined
by ${\cal H}^\prime_0(x)$. For constant matter density, the
oscillation probability for the $\nu_e \to \nu_\mu$ channel is found
to be \cite{Kopp:2007ne,Ribeiro:2007ud,Kikuchi:2008vq}
\begin{eqnarray}
P^{\rm NSI}_{e\mu} \simeq   P^{\rm SD}_{e\mu} &-& 4 \tilde s_{13}
s_{23} c_{23} \left( |\varepsilon_{e\mu}| c_{23} c_{\chi}-
|\varepsilon_{e\tau}| s_{23} c_{\omega} \right)
\left[\sin^2\frac{A\Delta}{2} - \sin^2\frac{\Delta}{2}
+\sin^2\frac{(1-A)\Delta}{2}  \right] \nonumber \\
&+&  8\tilde s_{13}s^2_{23} \left[ |\varepsilon_{e\mu}| s_{23}
c_{\chi} + |\varepsilon_{e\tau}| c_{23} c_{\omega}
\right]\frac{A}{1-A} \sin^2\frac{(1-A)\Delta}{2} \nonumber
\\
&+& 8 \tilde s_{13} s_{23} c_{23} \left( |\varepsilon_{e\mu}| c_{23}
s_{\chi} -|\varepsilon_{e\tau}| s_{23} s_{\omega} \right)
\sin\frac{A\Delta}{2}\sin\frac{\Delta}{2}\sin\frac{(1-A)\Delta}{2}
\; ,
\end{eqnarray}
where $\Delta \equiv \Delta^{}_{31} L$ denotes the oscillation phase
in vacuum, and $\tilde s_{13} = s_{13}/(1-A)$ is just $\sin
\tilde{\theta}_{13}$ in the leading order. In addition, we have
defined $\varepsilon^{}_{\alpha \beta} = |\varepsilon^{}_{\alpha
\beta}| e^{{\rm i}\phi^{}_{\alpha \beta}}$ (for $\alpha \beta =
e\mu, e\tau$), $\chi = \phi^{}_{e \mu} + \delta$, and $\omega =
\phi^{}_{e \tau} + \delta$. Note that we have neglected the terms
proportional to $\alpha~\varepsilon^{}_{\alpha \beta}$, and $P^{\rm
SD}_{e\mu}$ stands for the transition probability without the NSIs,
i.e, $P^{\rm SD}_{e\mu} \simeq 4\tilde s^2_{13} s^2_{23}
\sin^2\frac{(1-A)\Delta}{2}$. The approximate formulas of neutrino
oscillation probabilities to the second order of $\alpha$ and
$s_{13}$ can be found in Ref.~\cite{Akhmedov:2004ny}. At leading
order, only the NSI parameters $\varepsilon_{e\mu}$ and
$\varepsilon_{e\tau}$ appear in the transition probability $P^{\rm
NSI}_{e\mu}$ \cite{Kopp:2007ne, Ribeiro:2007ud}. Hence, we will
concentrate on these two parameters in following numerical analysis.
Furthermore, the CP-violating terms in the last line of Eq.~(21)
related to NSI parameters are not suppressed by the ratio of two
mass-squared differences $\alpha$, compared to the standard case.
Thus, even when the standard CP violation is not visible in an
experimental setup, one may expect observable CP-violating effects
stemming from the catalysis of NSIs.

Next, for the $\nu_\mu \to \nu_\mu$ channel, we have
\cite{Kopp:2007ne,Ribeiro:2007ud,Kikuchi:2008vq}
\begin{eqnarray}
P^{\rm NSI}_{\mu\mu} \simeq  P^{\rm SD}_{\mu\mu} &-&
|\varepsilon_{\mu\tau}| c_{\phi_{\mu\tau}} \left( s^3_{2\times 23}
A\Delta \sin\Delta + 4s_{2\times 23}c^2_{2\times
23}A\sin^2\frac{\Delta}{2}\right)
\nonumber \\
&+&  (|\varepsilon_{\mu\mu}| - |\varepsilon_{\tau\tau})|s^2_{2\times
23} c_{2\times 23} \left( \frac{A\Delta}{2} \sin\Delta -2
A\sin^2\frac{\Delta}{2}\right) \; ,
\end{eqnarray}
where $\varepsilon_{\mu \tau} \equiv |\varepsilon_{\mu \tau}|e^{{\rm
i}\phi_{\mu\tau}}$, $s^{}_{2\times 23} \equiv \sin 2\theta^{}_{23}$
and $c^{}_{2\times 23} \equiv \cos 2\theta^{}_{23}$ have been
defined. Note that only the NSI parameters $\varepsilon_{\mu\mu}$,
$\varepsilon_{\mu\tau}$, and $\varepsilon_{\tau\tau}$ appear in the
survival probability $P^{\rm NSI}_{\mu\mu}$ \cite{Kopp:2007ne}.
Since the current experimental bound on $\varepsilon_{\mu\mu}$ is
very stringent, the dominant NSI effects should come from
$\varepsilon_{\mu\tau}$ and $\varepsilon_{\tau\tau}$.

\section{Numerical Analysis}

In order to illustrate the NSI effects on neutrino propagation in
the Earth, we calculate numerically the effective mixing angle
$\tilde{\theta}_{13}$ and the oscillation probabilities. In our
numerical computations, we assume the Preliminary Reference Earth
model of Earth's matter density \cite{Dziewonski:1981xy}, and
express the baseline in terms of the zenith angle as $ L = - 2 R
\cos \theta_z$, where $R = 6371~{\rm km}$ is the Earth radius and
$\theta_z = \pi - h$ with $h$ being the nadir angle. Note that $-1 <
\cos \theta_z < -0.84$ corresponds to the trajectories crossing both
the mantle and core of the Earth, while $-0.84 < \cos \theta_z < 0$
to those crossing only the Earth mantle. On the other hand, there
already exist restrictive experimental constraints on the NSI
parameters in realistic models \cite{Davidson:2003ha}. However, in
Ref.~\cite{Biggio:2009nt}, the model-independent upper bounds on the
matter NSI parameters have been found to be much larger than the
model-dependent ones:
\begin{eqnarray}
\left(\begin{array}{lll}
|\varepsilon_{ee}| < 4.2 &
|\varepsilon_{e\mu}| < 0.33 &
|\varepsilon_{e\tau}| < 3.0 \\
~ & |\varepsilon_{\mu\mu}| < 0.068 &
|\varepsilon_{\mu\tau}| < 0.33 \\
~ & ~ & |\varepsilon_{\tau\tau}| < 21
\end{array}\right) \; .
\end{eqnarray}
Therefore, in the following discussions, we just ignore
$\varepsilon_{\mu\mu}$, which receives the most stringent
constraint. For the other matter NSI parameters, we will take a
conservative value $|\varepsilon_{\alpha \beta}| = 0.1$ for
illustration \footnote{Recently, the MINOS experiment has
constrained the NSI parameter to the range $-0.20 <
\varepsilon_{\mu\tau} < 0.07$ at the $90\%$ confidence level
\cite{MINOS}, in the framework of two-flavor neutrino
oscillations.}.

\subsection{Effective Mixing Angle in Matter}
\begin{figure}[]
\includegraphics[width=.48\textwidth]{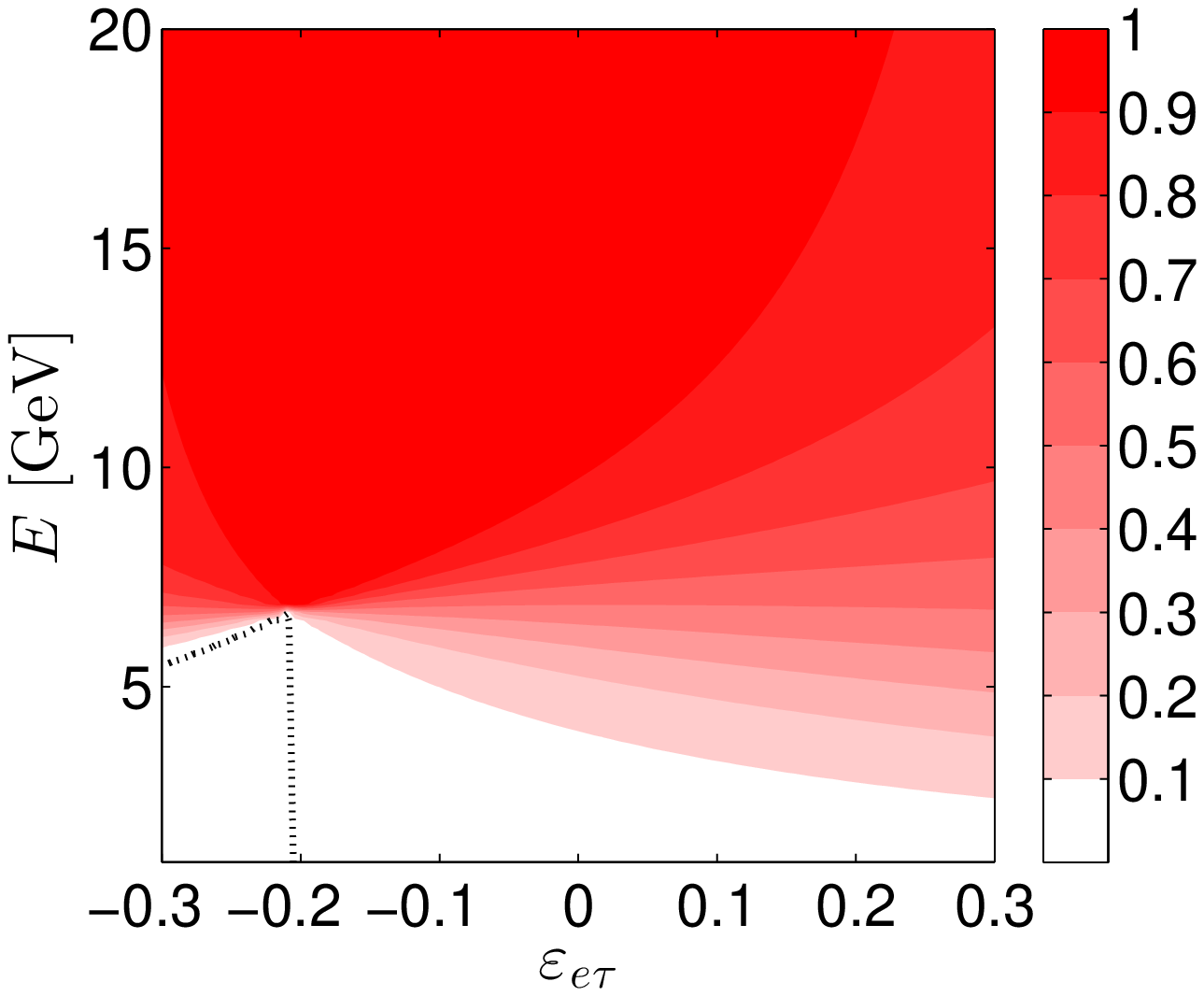}
\includegraphics[width=.48\textwidth]{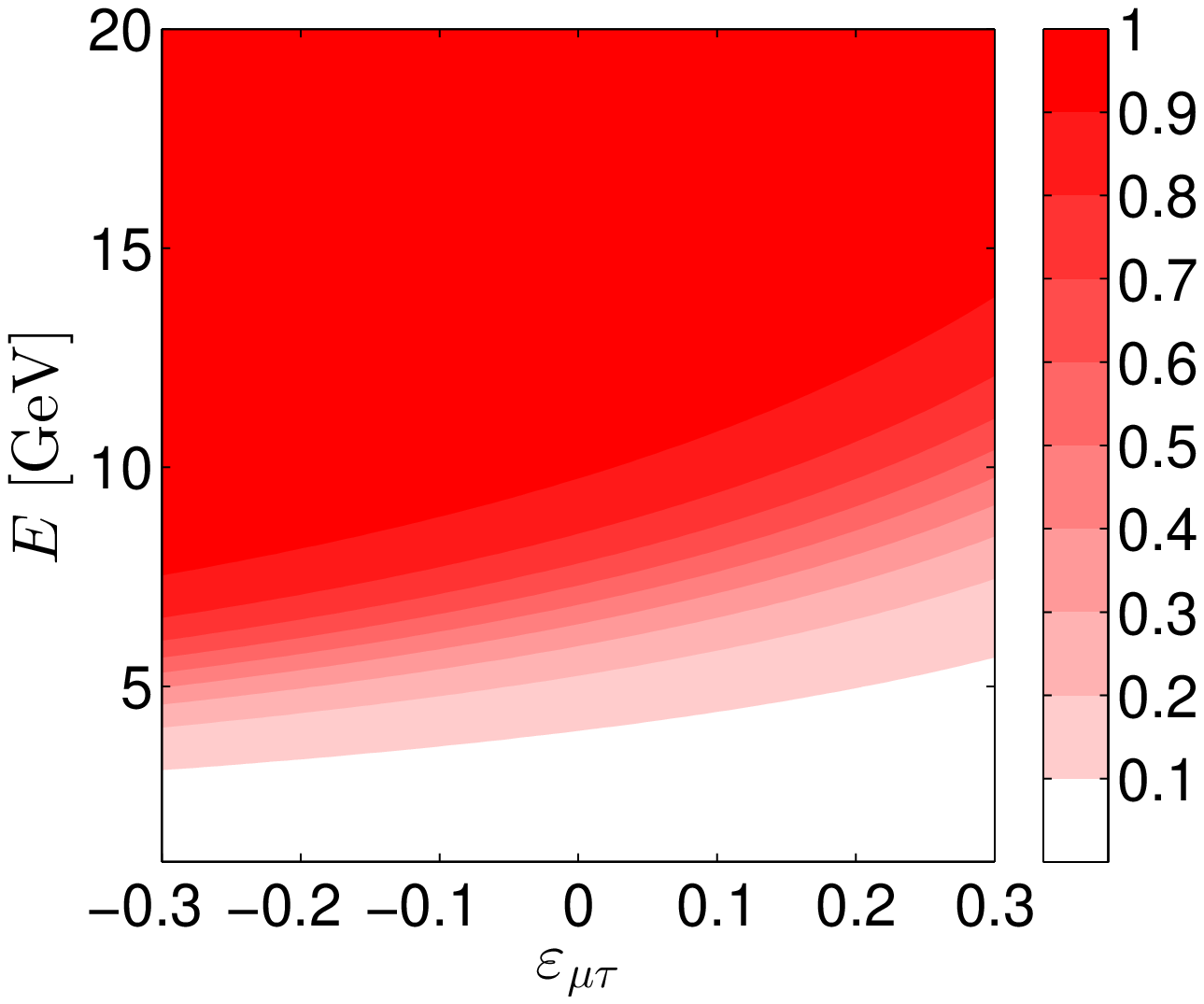}
\caption{\label{fig:theta} Dependence of the effective mixing angle
$\sin^2 \tilde{\theta}_{13}$ on the NSI parameters and neutrino
energy, where a constant matter density profile $\rho = 4.5~{\rm
g}/{\rm cm}^3$ has been assumed (i.e., the Earth mantle density) and
the dotted line corresponds to $\sin^2 \tilde{\theta}_{13} = \sin^2
\theta_{13}$. In addition, the best-fit values of neutrino
parameters (i.e., $\sin^2 \theta^{}_{12} = 0.30$, $\sin^2
\theta^{}_{23} = 0.41$, $\sin^2 \theta^{}_{13} = 0.023$, $\Delta
m^2_{21} = 7.50 \times 10^{-5}~{\rm eV}^2$, and $\Delta m^2_{31} =
2.47\times 10^{-3}~{\rm eV}^2$) from
Ref.~\cite{GonzalezGarcia:2012sz} have been used.}
\end{figure}
First of all, it may be interesting to show how the standard and
non-standard matter effects modify the effective neutrino mixing
angles $\tilde{\theta}_{ij}$ in matter. As mentioned in the previous
section, we will focus on $\tilde{\theta}_{13}$, which is relevant
for neutrino energies $E > 1~{\rm GeV}$. To examine the dependence
of $\tilde{\theta}_{13}$ on the NSI parameters, we consider two
specific examples, where only one relevant NSI parameter is switched
on in each case and all the CP-violating phases (i.e., $\delta$ and
$\phi_{\alpha \beta}$) are set to zero:
\begin{itemize}
\item $\varepsilon_{e\tau} \neq 0$. With the
help of Eqs.~(9) and (19), we derive
\begin{equation}
\sin^2\tilde{\theta}_{13} = \frac{\hat{C} - \cos2\theta_{13} +
A}{2\hat{C}} + \frac{\cos2\theta_{13} - A}{\hat{C}^3} A
\sin2\theta_{13} c_{23} \varepsilon_{e\tau} \; ,
\end{equation}
where the small terms $\alpha s_{13}$ and $\varepsilon^2_{e\tau}$
have been neglected. It is worthwhile to mention that if the
resonance condition $A = \cos 2\theta_{13}$ is satisfied, we have
$\tilde{\theta}_{13} = \pi/4$, which is independent of the NSI
parameter $\varepsilon_{e\tau}$ in the leading order approximation.
This can be well understood in the framework of two-flavor neutrino
oscillations in matter with NSIs
\cite{Blennow:2005qj,Blennow:2008eb}, where one observes that the
off-diagonal term in ${\cal H}_{\rm NSI}$ cannot modify the
resonance condition. The result in the case of $\varepsilon_{e\mu}
\neq 0$ can be obtained by replacing $c_{23} \varepsilon_{e\tau}$
with $s_{23} \varepsilon_{e\mu}$ in Eq~(24), and the difference
between these two cases can be attributed to a non-maximal
$\theta_{23}$. In Fig.~\ref{fig:theta}, we have calculated
$\sin^2\tilde{\theta}_{13}$ using the exact formulas of parameter
mappings \cite{Meloni:2009ia}. From the left panel, we can clearly
observe that the resonance condition is essentially unchanged by
$\varepsilon_{e\tau}$, and $\tilde{\theta}_{13} = \theta_{13}$ is
achieved when the standard matter effects are cancelled by the NSI
effects. However, the resonance is in fact shifted by higher-order
corrections.

\item $\varepsilon_{\mu\tau} \neq 0$. In a similar way, we obtain
\begin{equation}
\sin^2\tilde{\theta}_{13} = \frac{\hat{C} - \cos2\theta_{13} +
A}{2\hat{C}} - \frac{\sin^2 2\theta_{13}}{2\hat{C}^3} A \sin
2\theta_{23} \varepsilon_{\mu\tau} \; ,
\end{equation}
where the small terms $\alpha s_{13}$ and $\varepsilon^2_{\mu\tau}$
have been omitted. It is now evident that the standard resonance
condition $A = \cos 2\theta_{13}$ does not lead to $\sin^2
\tilde{\theta}_{13} = 1/2$. Namely, the resonance has been shifted
by the NSI parameter, which has also been pointed out in
Ref.~\cite{Blennow:2005qj} in the framework of two-neutrino
oscillations. For a fixed value of $\tilde{\theta}_{13}$, if
$\varepsilon_{\mu\tau}$ becomes larger, the neutrino energy has to
increase in order to balance the negative contribution from
$\varepsilon_{\mu\tau}$. In the right panel of Fig.~\ref{fig:theta},
we have shown $\sin^2\tilde{\theta}_{13}$ by using the exact mapping
formulas. The main features can be well described by the approximate
formula in Eq.~(25). Note that, compared to $\varepsilon_{e\tau}$,
the $\varepsilon_{\mu\tau}$ correction to $\tilde \theta_{13}$ is
milder since it is further suppressed by $\sin2\theta_{13}$.
Similarly, one can also consider the impact of $\varepsilon_{ee}$,
$\varepsilon_{\mu\mu}$, and $\varepsilon_{\tau\tau}$ on the
effective mixing angles.
\end{itemize}

\subsection{Neutrino Oscillograms: Standard Case}
\begin{figure}[]
\includegraphics[width=.48\textwidth]{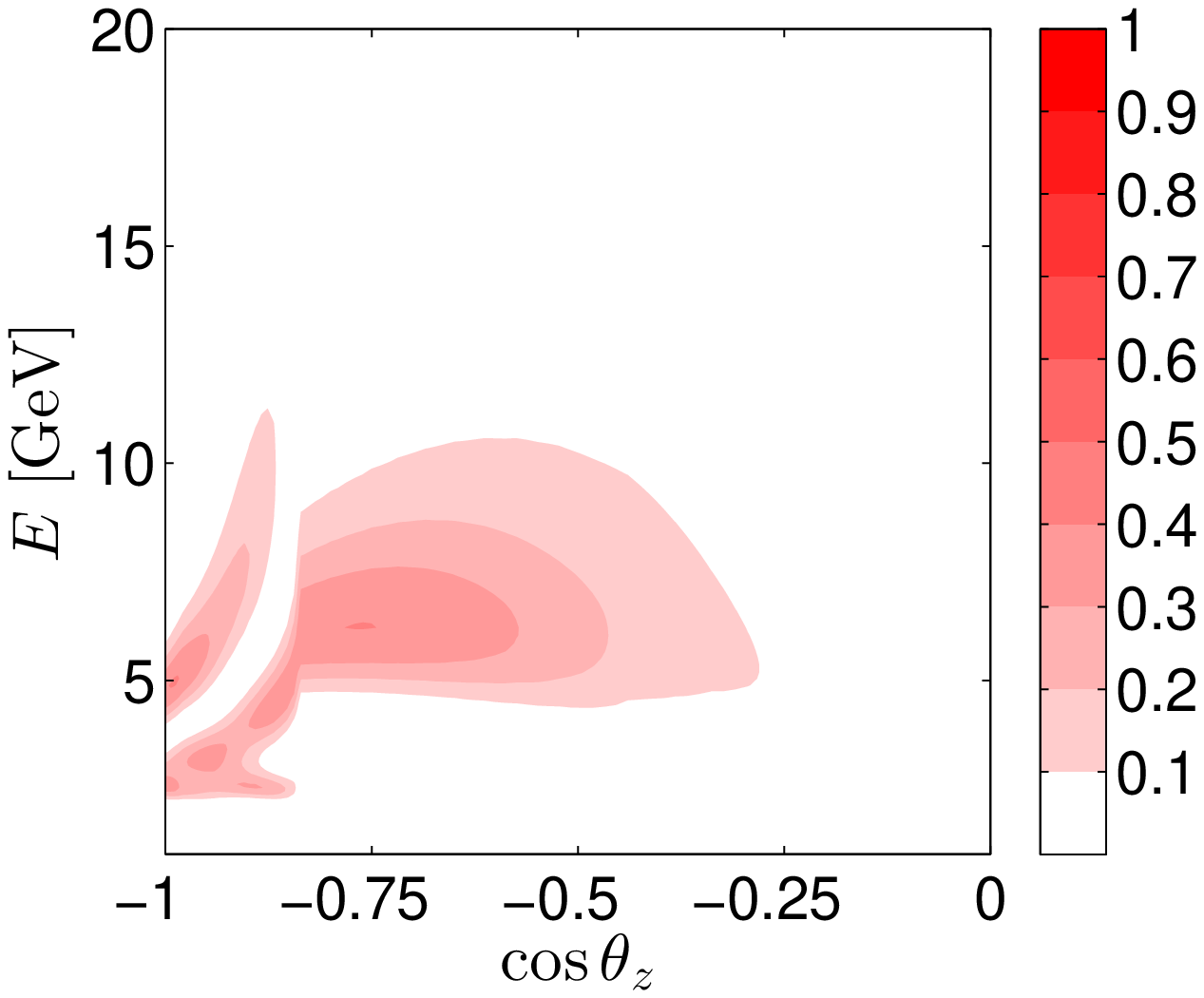}
\includegraphics[width=.48\textwidth]{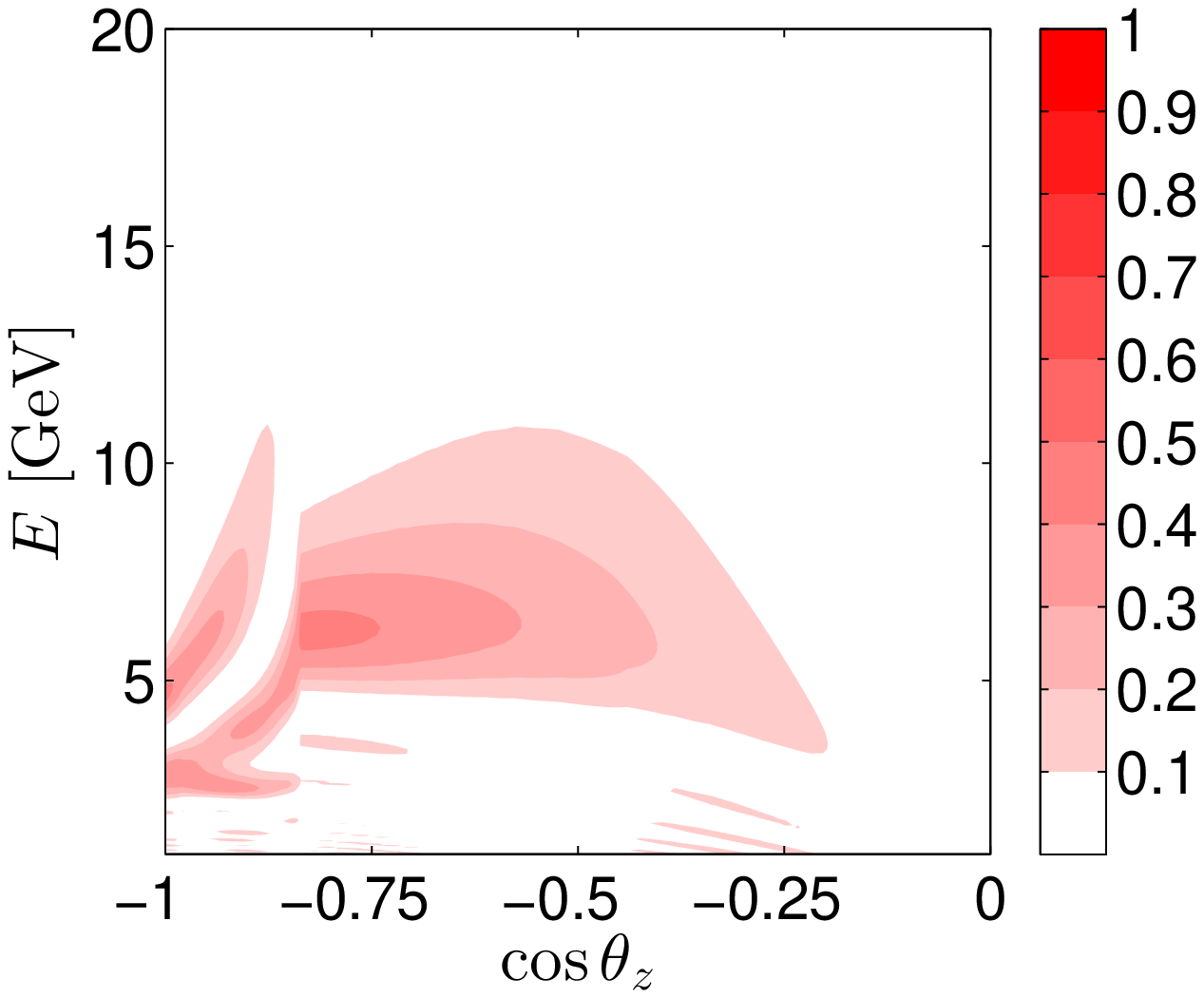}
\caption{\label{fig:P1} Standard neutrino oscillograms without NSIs
(i.e., $\varepsilon_{\alpha \beta} = 0$) in the appearance channel:
$P^{\rm SD}_{e\mu} = P(\nu_e \to \nu_\mu)$ for neutrino oscillations
in the case of normal neutrino mass hierarchy (left panel) and
$\bar{P}^{\rm SD}_{e\mu} = P(\bar{\nu}_e \to \bar{\nu}_\mu)$ for
antineutrino oscillations in the case of inverted neutrino mass
hierarchy (right panel).}
\end{figure}

\begin{figure}[]
\includegraphics[width=.48\textwidth]{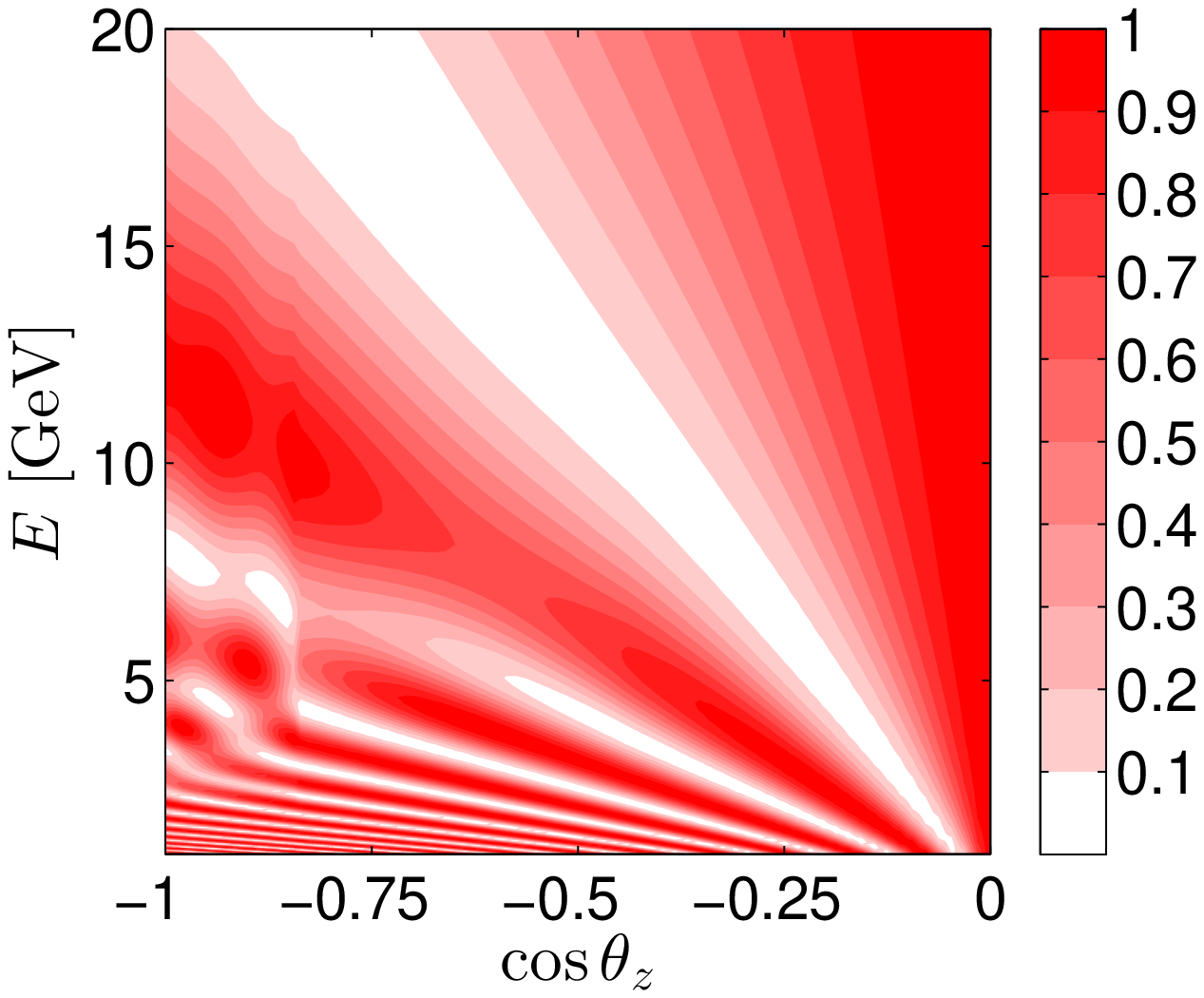}
\includegraphics[width=.48\textwidth]{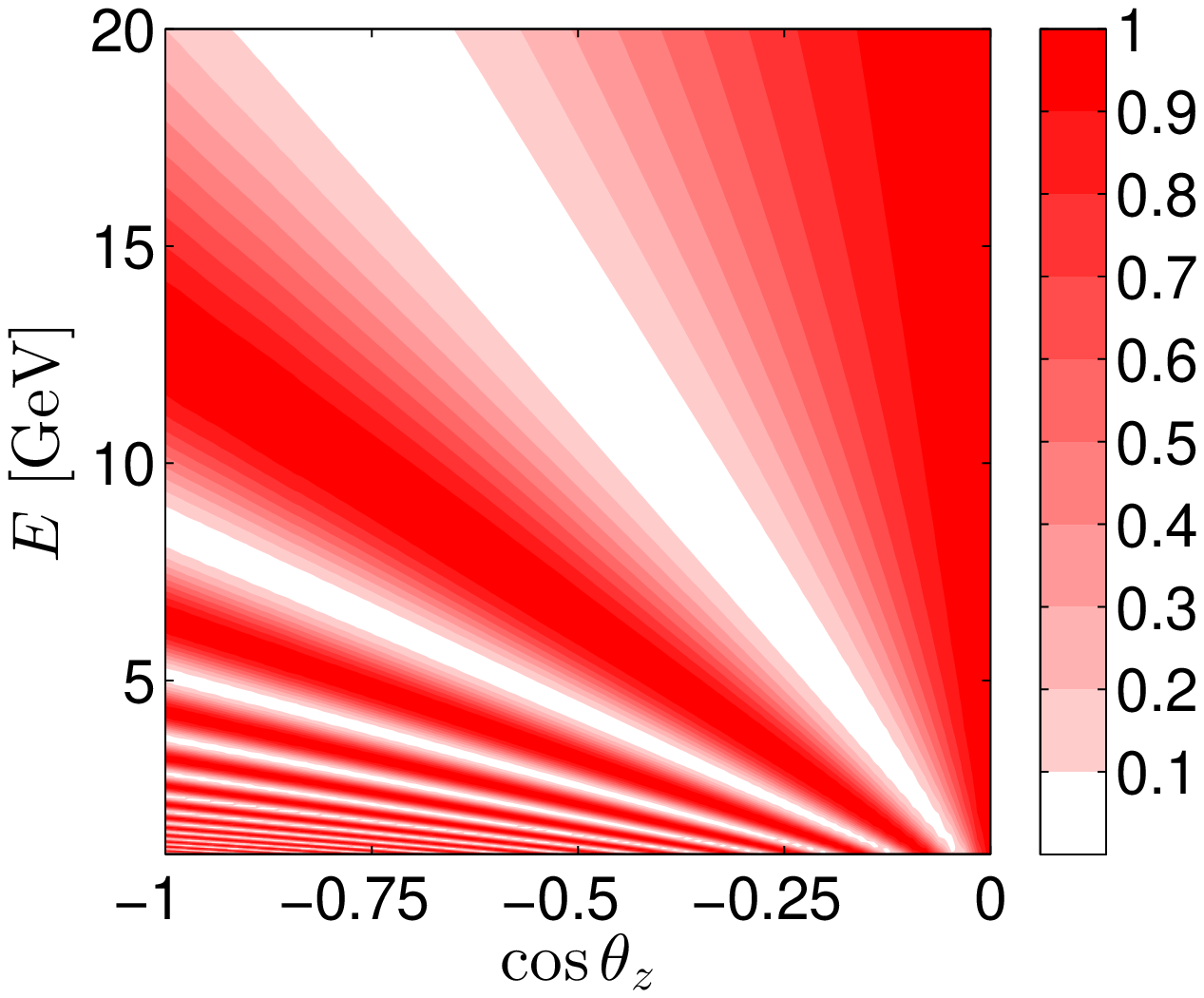}
\caption{\label{fig:P2} Standard neutrino oscillograms without NSIs
(i.e., $\varepsilon_{\alpha \beta} = 0$) in the disappearance
channel: $P^{\rm SD}_{\mu\mu} = P(\nu_\mu \to \nu_\mu)$ for neutrino
oscillations in the case of normal neutrino mass hierarchy (left
panel) and in the case of inverted neutrino mass hierarchy (right
panel).}
\end{figure}
The matter effects on neutrino propagation in the Earth can be
perfectly illustrated through the so-called neutrino oscillograms.
In order to compare between standard and non-standard matter
effects, we first briefly summarize the general features of the
standard neutrino oscillograms, which have been systematically
studied in Refs.~\cite{Akhmedov:2006hb,Akhmedov:2008qt}.

In Figs.~\ref{fig:P1} and \ref{fig:P2}, we have reproduced the
neutrino oscillograms in the $\nu_e \to \nu_\mu$ and $\nu_\mu \to
\nu_\mu$ channels, respectively. In our numerical calculations, the
latest global-fit data on leptonic mixing angles (i.e., $\sin^2
\theta^{}_{12} = 0.30$, $\sin^2 \theta^{}_{23} = 0.41$, and $\sin^2
\theta^{}_{13} = 0.023$) and the neutrino mass-squared differences
(i.e., $\Delta m^2_{21} = 7.50 \times 10^{-5}~{\rm eV}^2$ and
$|\Delta m^2_{31}| = 2.47\times 10^{-3}~{\rm eV}^2$) given in
Ref.~\cite{GonzalezGarcia:2012sz} have been used. Since the Dirac
CP-violating phase has not been experimentally constrained, we
simply take $\delta=0$ in the calculations. Some general comments on
the neutrino oscillograms are in order~\cite{Akhmedov:2006hb}:
\begin{itemize}
\item In the left panel of Fig.~\ref{fig:P1}, the oscillation probability
$P^{\rm SD}_{e\mu} \equiv P(\nu_e \to \nu_\mu)$ is shown in the NH
case. The resonance in the mantle appears around $\cos \theta^{\rm
R}_z \approx -0.75$ and $E_{\rm R} \approx 6~{\rm GeV}$, where the
resonance energy $E_{\rm R}$ is essentially determined by $A = \cos
2\theta_{13}$ while the corresponding baseline $L_{\rm R}$ or the
zenith angle $\theta^{\rm R}_z$ by the requirement of maximal
oscillation phase $\Delta = \pi$. Note that we can safely ignore the
effects of $\Delta m^2_{21}$ for neutrino energies $E > 1~{\rm
GeV}$. The ridges in the core region are caused by the parametric
resonances, receiving both contributions from the mantle and core
oscillation phases \cite{Akhmedov:2006hb}. Since the resonance takes
place in the neutrino sector, the oscillation probability of
$\bar{\nu}_e \to \bar{\nu}_\mu$ is suppressed by matter effects and
the oscillogram is almost empty. In the right panel of
Fig.~\ref{fig:P1}, we have calculated the antineutrino oscillation
probability $\bar{P}^{\rm SD}_{e\mu} \equiv P(\bar{\nu}_e \to
\bar{\nu}_\mu)$ in the IH case. Now that the resonances occur in
this case and dominate the contributions to oscillation
probabilities, the similarity between $P^{\rm SD}_{e\mu}$ for NH and
$\bar{P}^{\rm SD}_{e\mu}$ for IH is evident. The oscillation
probability $\bar{P}^{\rm SD}_{e\mu}$ in the NH case is highly
suppressed, as $P^{\rm SD}_{e\mu}$ in the IH case.

\item In Fig.~\ref{fig:P2}, we have given the survival
probabilities $P^{\rm SD}_{\mu\mu} \equiv P(\nu_\mu \to \nu_\mu)$
for both NH (left panel) and IH (right panel). The probability
$P^{\rm SD}_{\mu\mu}$ receives the dominant contribution from the
vacuum oscillation due to $\theta_{23}$ and $\Delta m^2_{31}$, and
is significantly affected by the $1$-$3$ mixing through
$\theta_{13}$ only in the resonance regions. In fact, the latter
effect reduces $P^{\rm SD}_{\mu\mu}$, as indicated in the left panel
of Fig.~\ref{fig:P2}. In the absence of resonance, as in the IH
case, $P^{\rm SD}_{\mu\mu}$ is basically given by the vacuum
oscillation probability as shown in the right panel. The
antineutrino survival probabilities $\bar{P}^{\rm SD}_{\mu\mu}
\equiv P(\bar{\nu}_\mu \to \bar{\nu}_\mu)$ in NH and IH cases are
similar to $P^{\rm SD}_{\mu\mu}$ in IH and NH cases, respectively.
\end{itemize}

The detailed study of the maxima and minima in neutrino oscillation
probabilities and the general conditions for resonances in the
two-flavor approximation can be found in
Ref.~\cite{Akhmedov:2006hb}, while those for the three-flavor
oscillations in Ref.~\cite{Akhmedov:2008qt}. In our discussions,
since the neutrino energies are always above $1~{\rm GeV}$, the
three-flavor corrections should be negligible.

\subsection{Neutrino Oscillograms: NSI Effects}
\begin{figure}[!h]\vspace{-0.3cm}
\includegraphics[width=.48\textwidth]{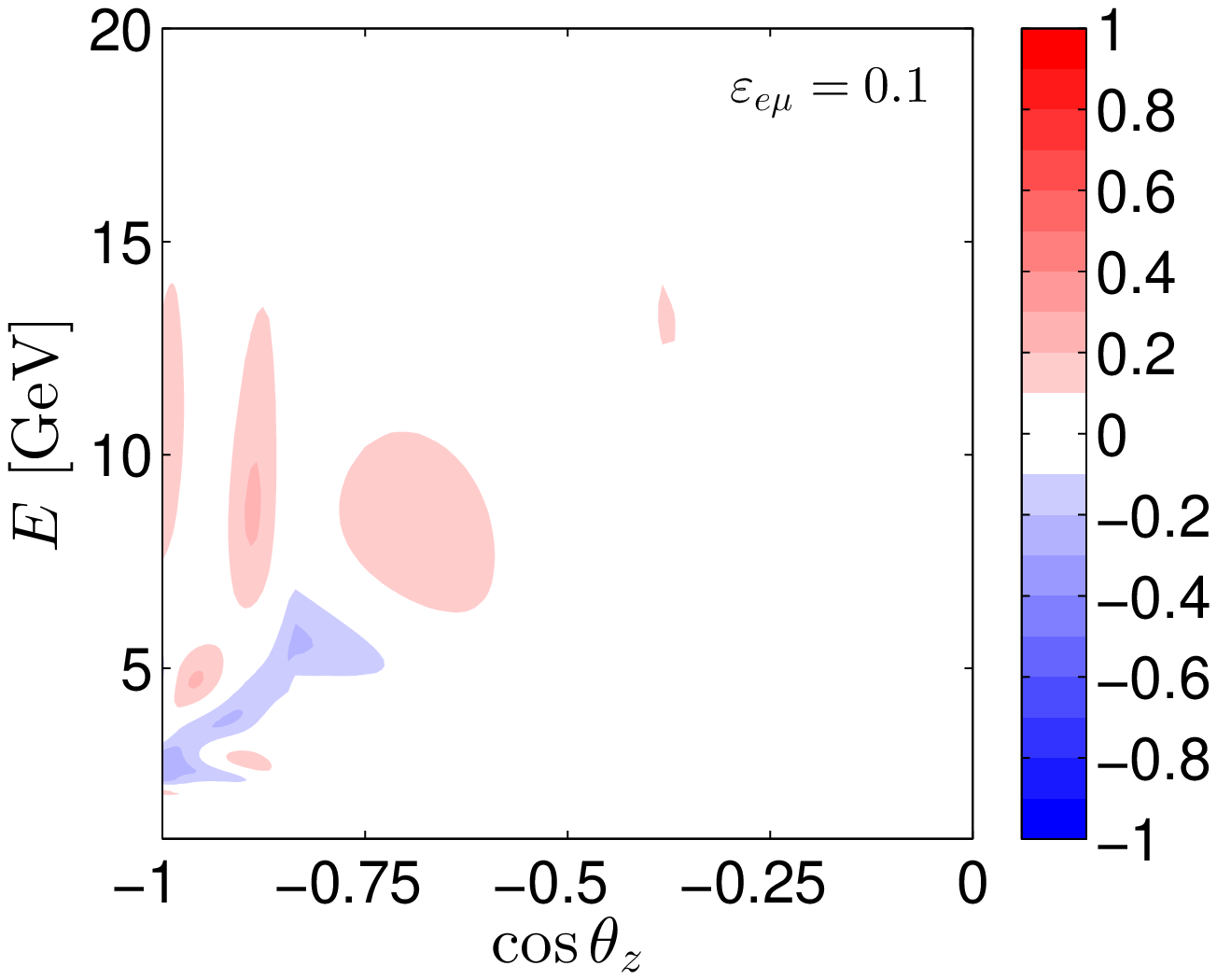}
\includegraphics[width=.48\textwidth]{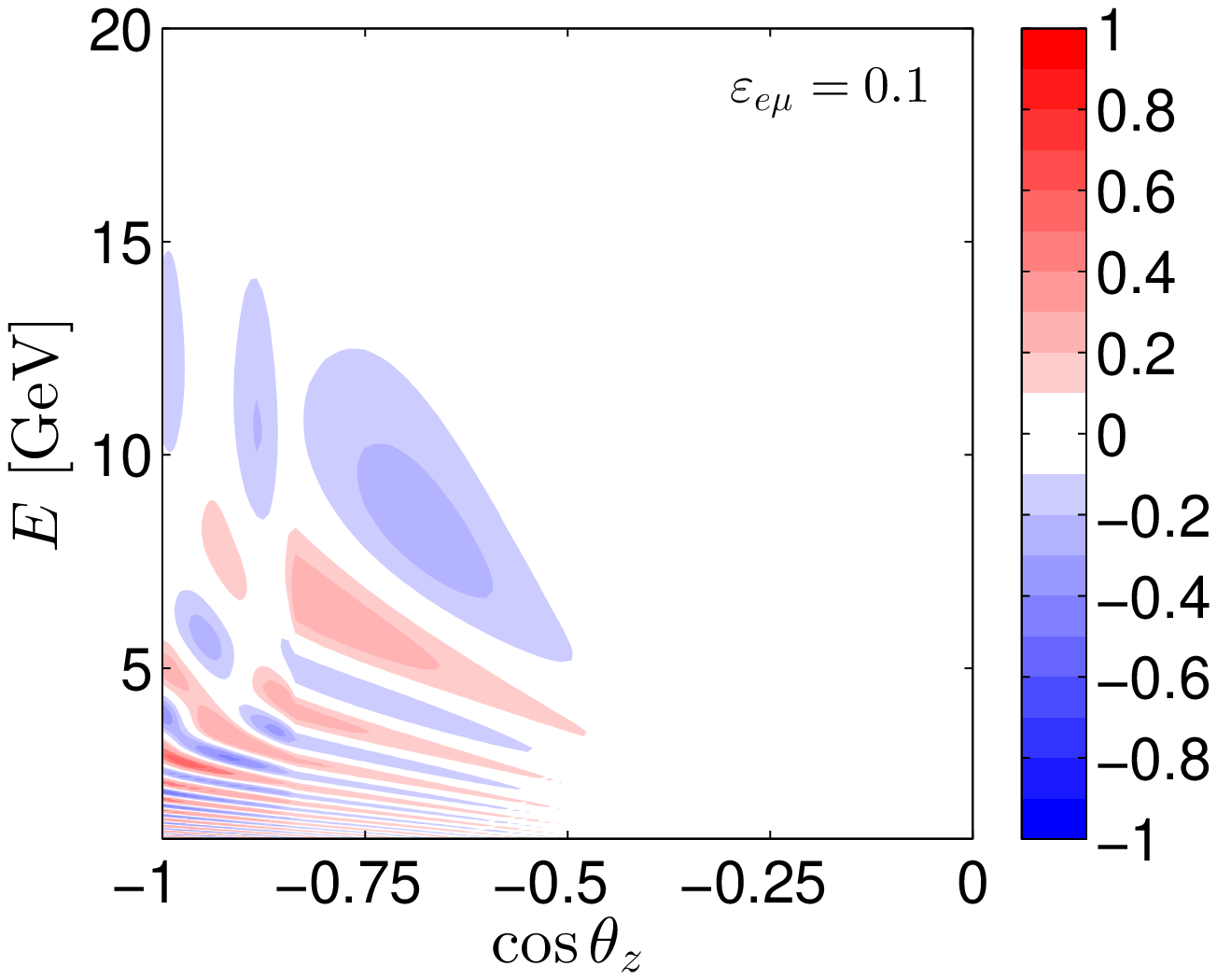}
\includegraphics[width=.48\textwidth]{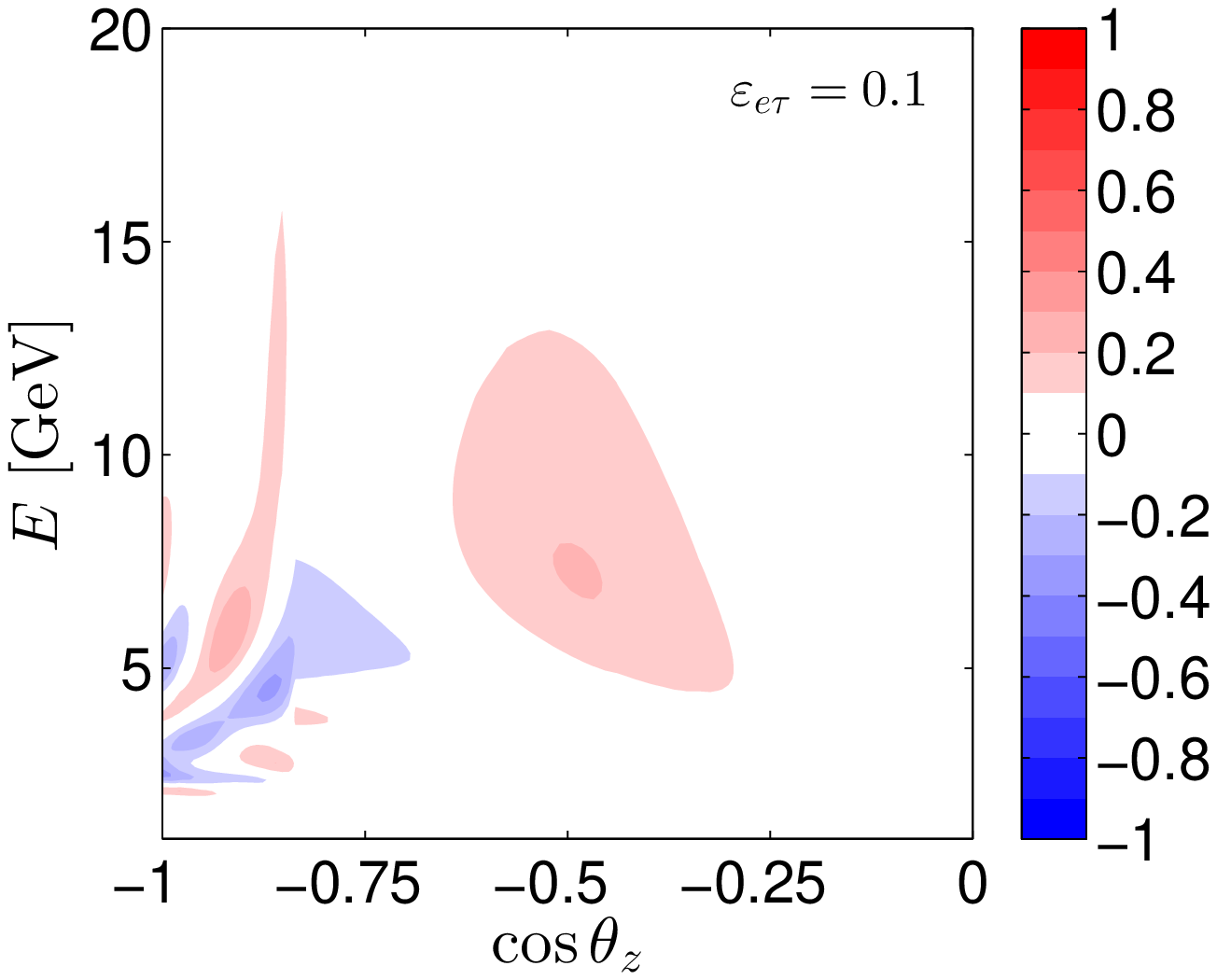}
\includegraphics[width=.48\textwidth]{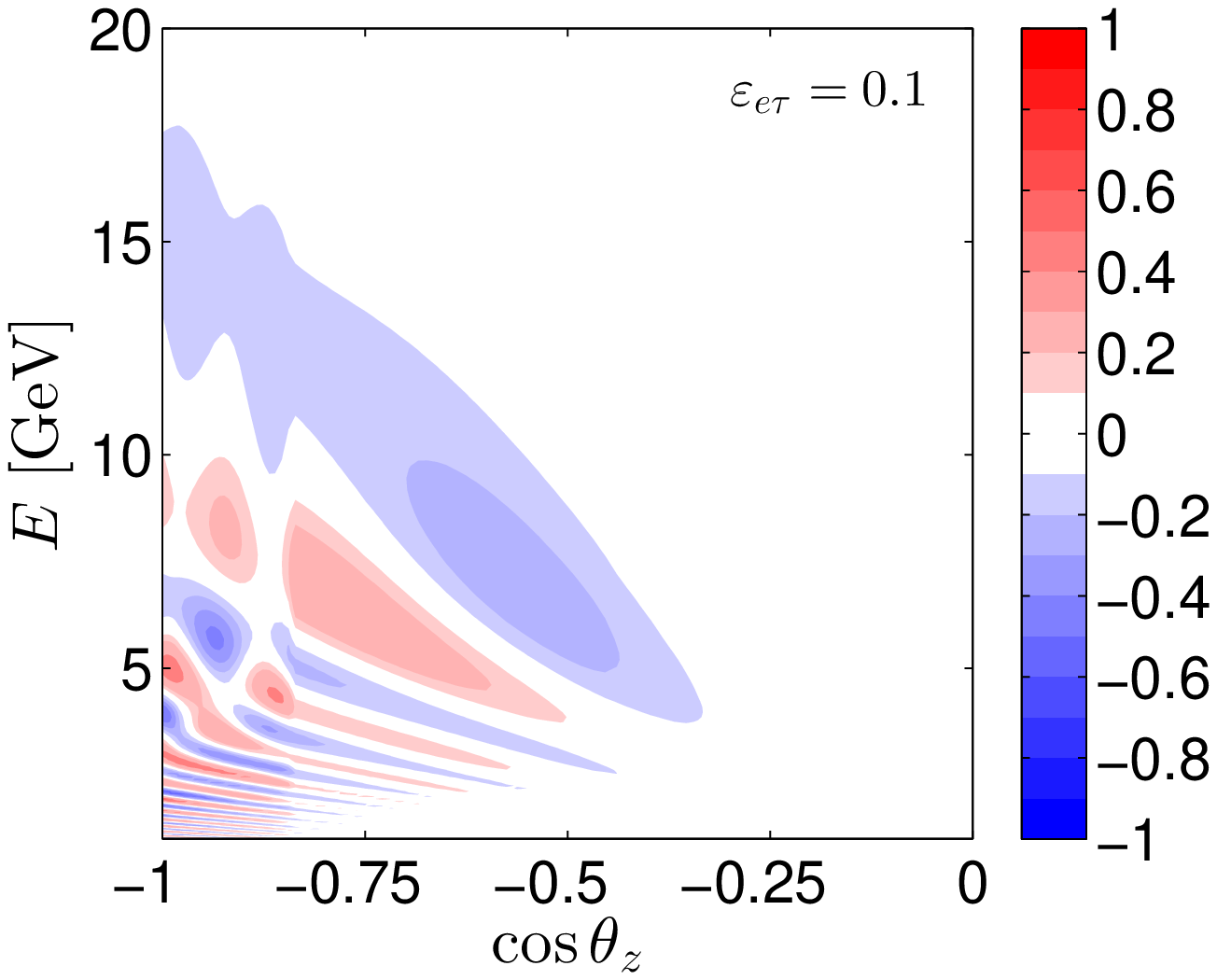}
\includegraphics[width=.48\textwidth]{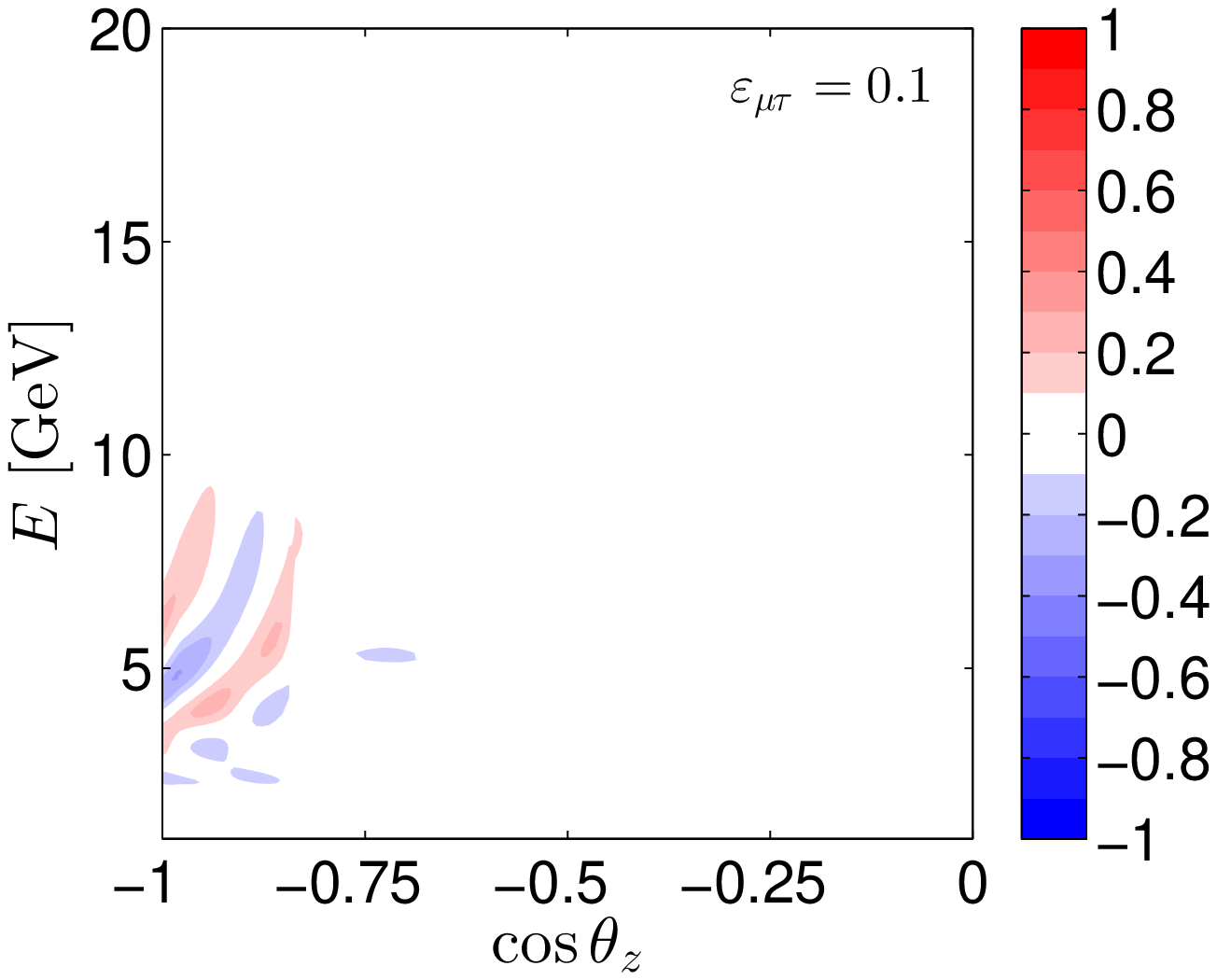}
\includegraphics[width=.48\textwidth]{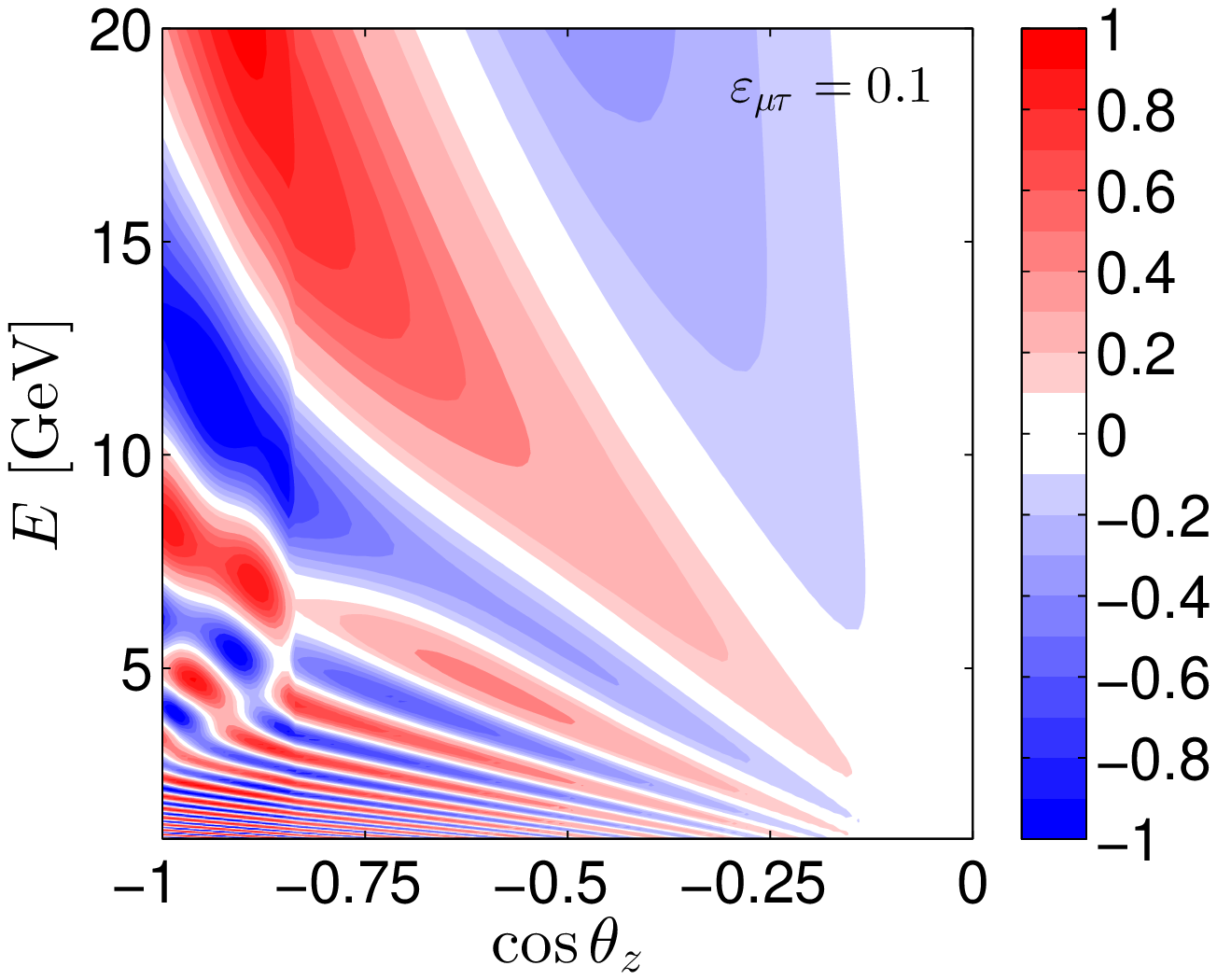}
\caption{\label{fig:DeltaP} Differences between the standard and
non-standard neutrino oscillograms in the $\nu_e \to \nu_\mu$
channel $\Delta P_{e\mu} \equiv P^{\rm NSI}_{e\mu} - P^{\rm
SD}_{e\mu}$ (left column) and in the $\nu_\mu \to \nu_\mu$ channel
$\Delta P_{\mu\mu} \equiv P^{\rm NSI}_{\mu\mu} - P^{\rm
SD}_{\mu\mu}$ (right column), where the normal neutrino mass
hierarchy and $\delta = 0$ are assumed, and $\varepsilon_{e\mu} =
0.1$ (upper row), $\varepsilon_{e\tau} = 0.1$ (middle row), and
$\varepsilon_{\mu\tau} = 0.1$ (lower row) are taken for
illustration.}
\end{figure}
Now we proceed to discuss the NSI effects on the standard neutrino
oscillations in the Earth by using neutrino oscillograms. To
quantify the NSI effects, we consider the difference between the
standard (i.e., $\varepsilon_{\alpha\beta} = 0$) and non-standard
(i.e., $\varepsilon_{\alpha\beta} \neq 0$) neutrino oscillograms,
namely $\Delta P_{e\mu} = P^{\rm NSI}_{e\mu} -P^{\rm SD}_{e\mu}$ and
$\Delta P_{\mu\mu} = P^{\rm NSI}_{\mu\mu} -P^{\rm SD}_{\mu\mu}$. In
Fig.~\ref{fig:DeltaP}, we numerically calculate $\Delta P_{e\mu}$
and $\Delta P_{\mu\mu}$ in the NH case, and only one NSI parameter
is switched on in each plot.

First, we consider the difference $\Delta P_{e\mu}$ in the
appearance channel, for which the approximate formula in the case of
constant matter density can be obtained from Eq.~(21) and the
numerical results are shown in the left column of
Fig.~\ref{fig:DeltaP}. Two comments are in order: (1) The
contributions from NSI parameters to $\Delta P_{e\mu}$ are
proportional to $s_{13}$ in the limit $A \to 1$, i.e., around the
resonance. In the deep core, i.e., $\cos \theta_z < -0.84$, the
parametric resonances dominate, so the NSI effects are not
necessarily suppressed. (2) The difference $\Delta P_{e\mu}$ is
independent of $\varepsilon_{\mu\tau}$ in leading order. For this
reason, as shown in the lower plot, $\Delta P_{e\mu}$ is vanishing
everywhere except in the core, if only $\varepsilon_{\mu\tau} \neq
0$ is assumed. One common feature of all three plots is that
significant NSI effects are lying in the core region, which is only
accessible in the atmospheric neutrino experiments.

Second, we turn to the difference $\Delta P_{\mu\mu}$ in the
disappearance channel, for which the approximate formula in the case
of constant matter density can be obtained from Eq.~(22) and the
numerical results are shown in the right column of
Fig.~\ref{fig:DeltaP}. As implied by Eq.~(22), the effects induced
by $\varepsilon_{e\mu}$ and $\varepsilon_{e\tau}$ can only arise
from higher-order corrections of $s_{13}$ and $\alpha$, so they are
insignificant as in the upper and middle plots. The most interesting
observation is that $|\Delta P_{\mu\mu}| \simeq 1$ could be
achieved, particularly in the core region. Switching off both
$\varepsilon_{\mu\mu}$ and $\varepsilon_{\tau\tau}$ in Eq.~(22), we
have
\begin{equation}
\Delta P_{\mu\mu} \approx - |\varepsilon_{\mu\tau}| \sin
2\theta_{23} \left(A\Delta \sin^2 2\theta_{23} \sin \Delta + 4\cos^2
2\theta_{23} \sin^2 \frac{\Delta}{2}\right) \; ,
\end{equation}
where $\phi_{\mu\tau} = 0$ is assumed. Furthermore, note that the
second term in the parentheses on the right-hand side of Eq.~(26) is
always positive, while the first term can be either positive or
negative, depending on the oscillation phase $\Delta$. If $\Delta =
(2k+1)\pi/2$ with $k$ being an integer, the contribution from the
second term is negligible, since $\cos^2 2\theta_{23} \approx 0.03$
for the best-fit value of $\theta_{23} = 40^\circ$. Therefore,
$\Delta P_{\mu\mu} \propto \sin \Delta$ should show an oscillatory
behavior, as given in the lower plot. It is worthwhile to note that
$\Delta = (2k+1)\pi/2$ corresponds to the oscillation minima of the
standard survival probability $P^{\rm SD}_{\mu\mu}$. For example,
comparing the left plot of Fig.~\ref{fig:P2} with the lower plot in
the right column of Fig.~\ref{fig:DeltaP}, one can observe the huge
difference in the region along the diagonal line. If $\Delta = k\pi$
holds, the second term dominates over the first one, and $\Delta
P_{\mu\mu}$ follow the same oscillatory structure as the leading
vacuum oscillation term in $P^{\rm SD}_{\mu\mu}$.

\subsection{Non-Standard CP Violation}
\begin{figure}[!h]\vspace{-0.3cm}
\includegraphics[width=.48\textwidth]{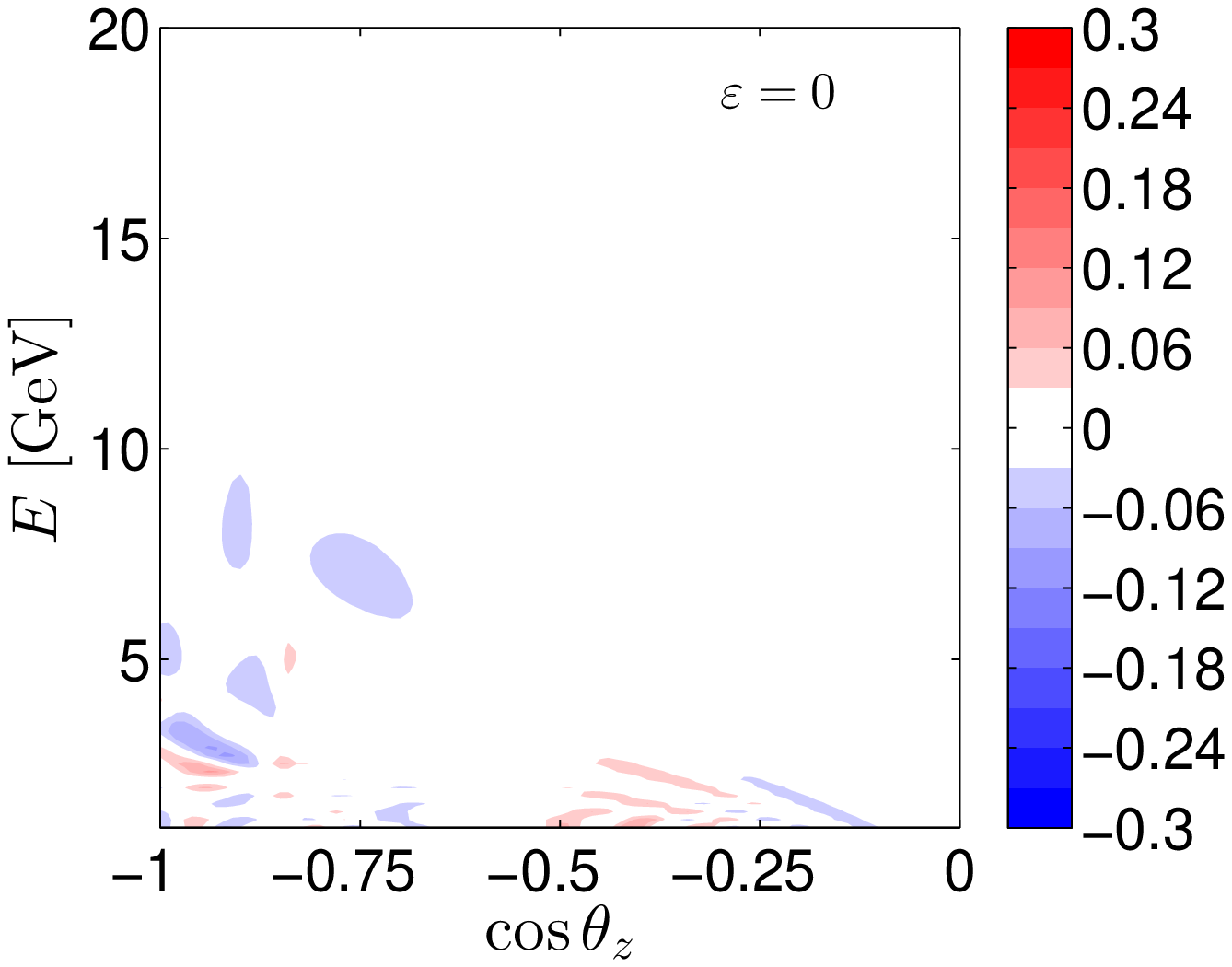}
\includegraphics[width=.48\textwidth]{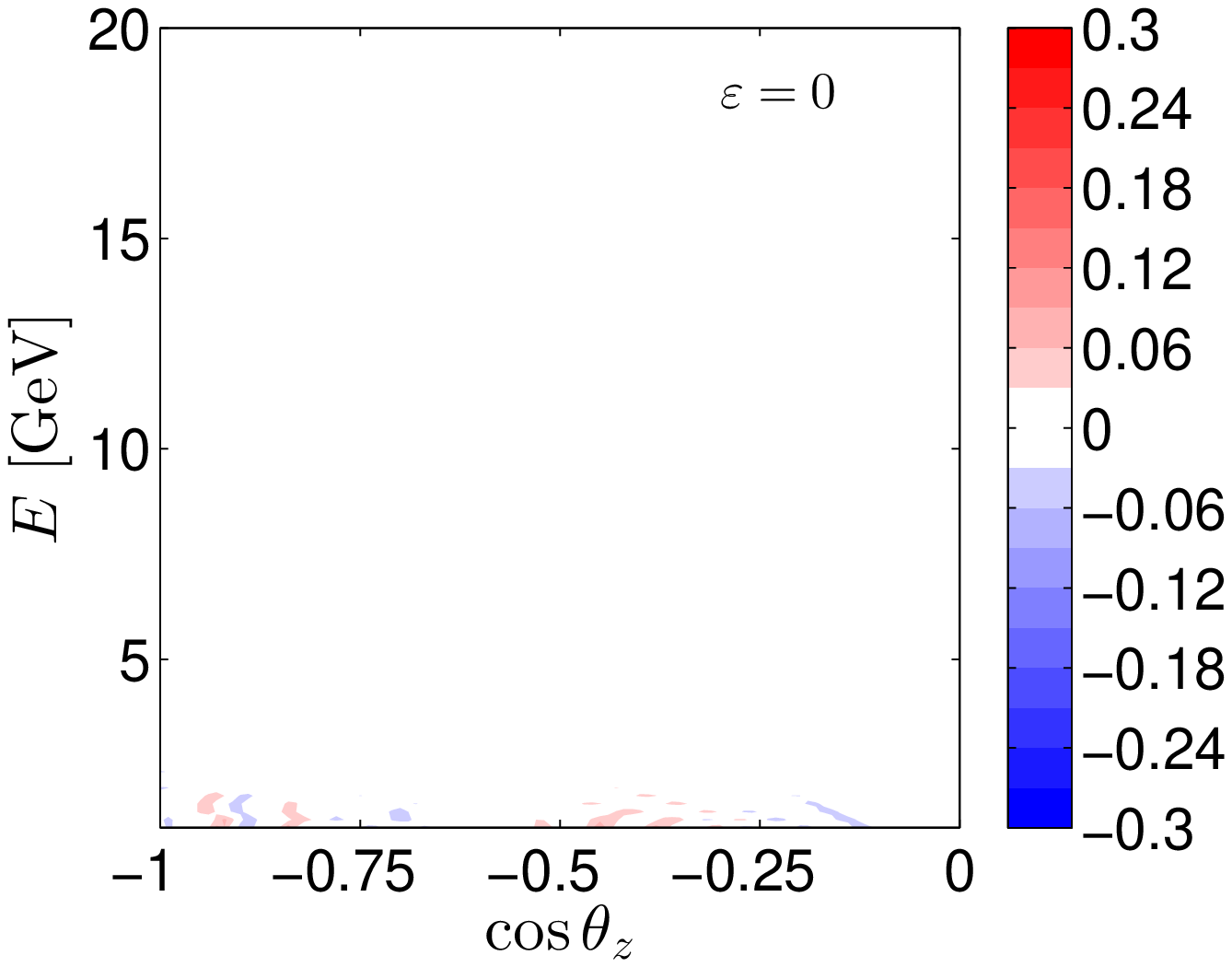}
\includegraphics[width=.48\textwidth]{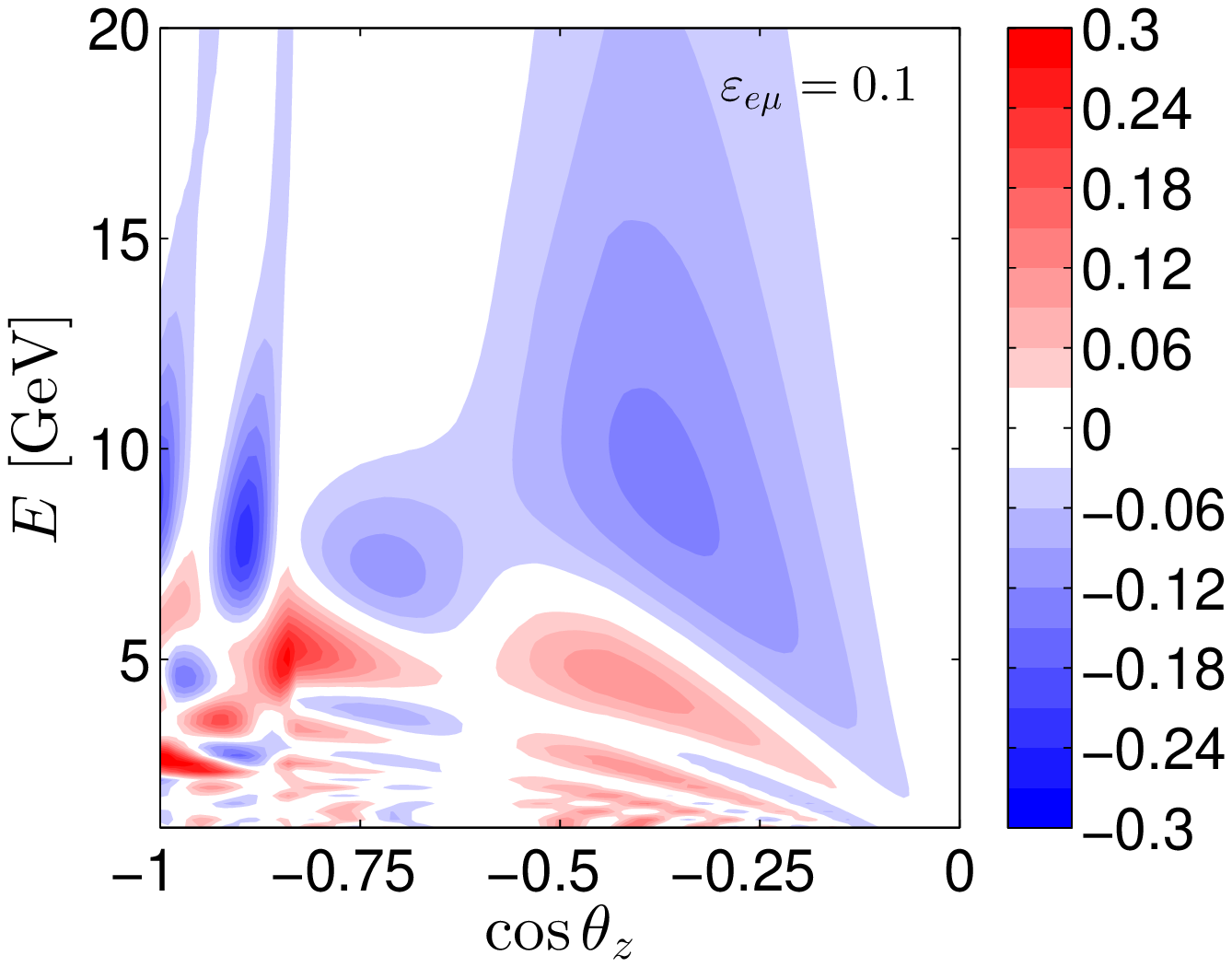}
\includegraphics[width=.48\textwidth]{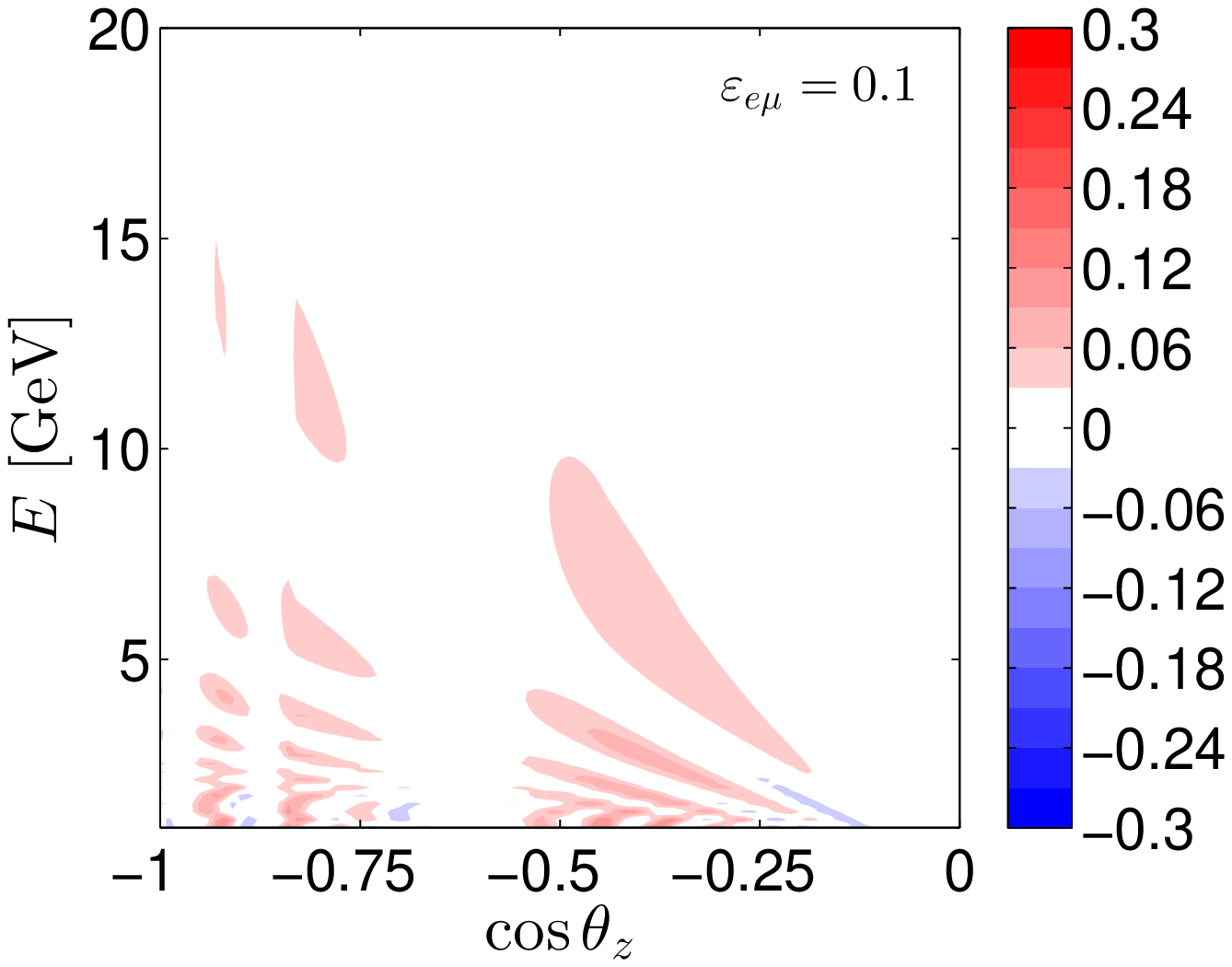}
\includegraphics[width=.48\textwidth]{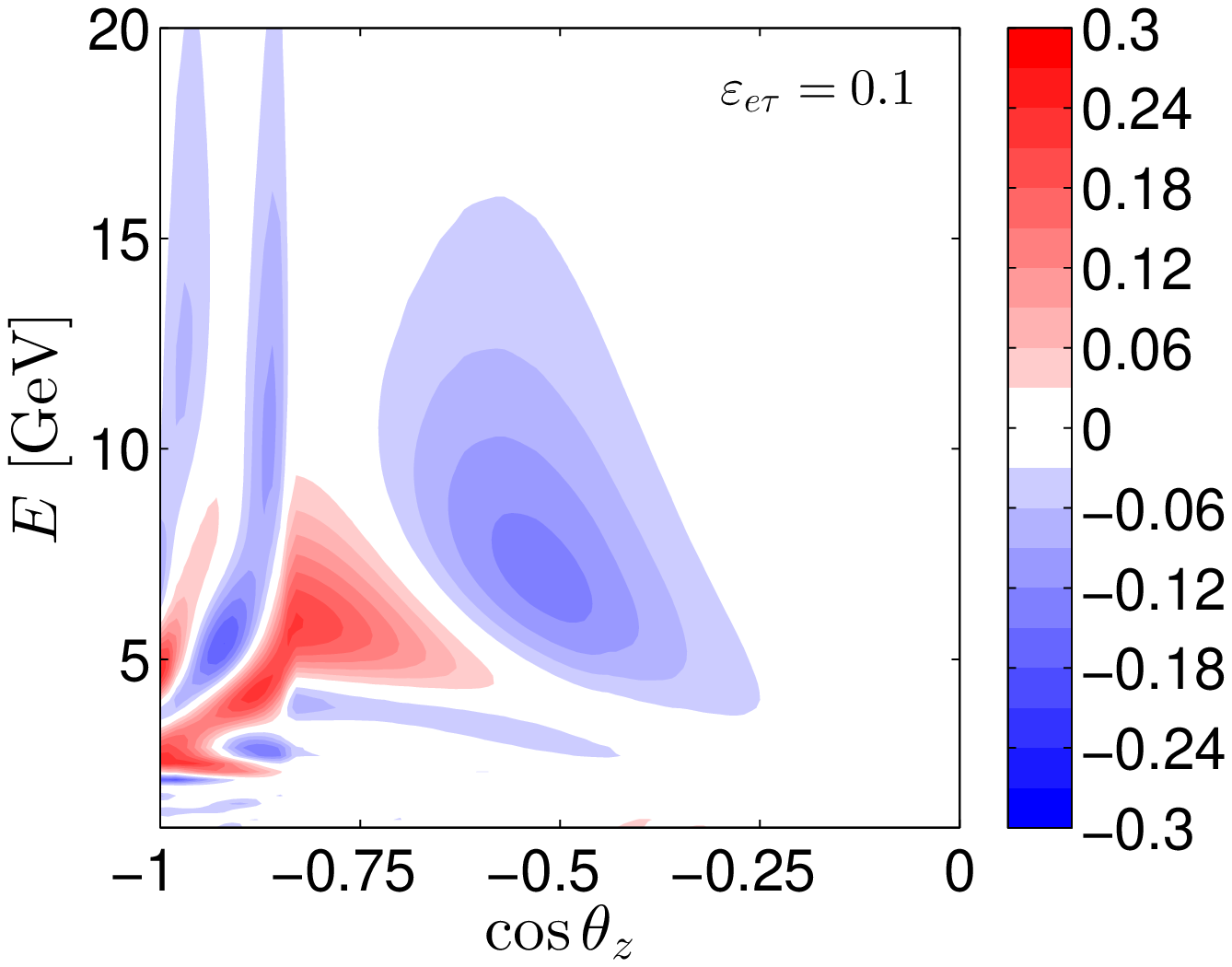}
\includegraphics[width=.48\textwidth]{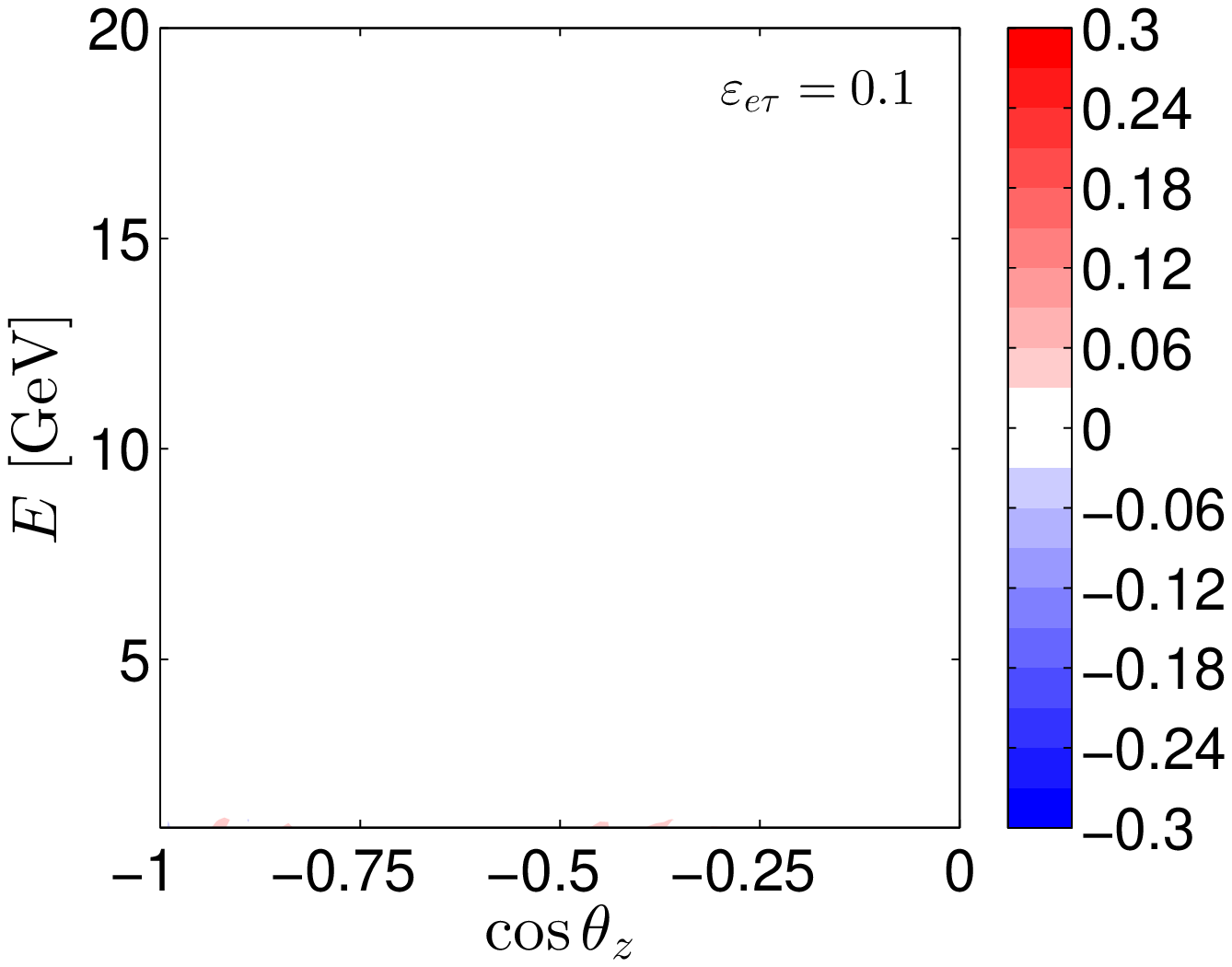}
\caption{\label{fig:CP} Oscillograms for the probability differences
$\Delta P_{e\mu}(\delta) = P^{\rm NSI}_{e\mu}(\delta) -P^{\rm
NSI}_{e\mu}(0)$ with $\delta=\pi/2$ in the normal neutrino mass
hierarchy case (left column) and in the inverted neutrino mass
hierarchy case (right column). The two plots in the first row
correspond to $\Delta P_{e\mu}(\delta)$ with  vanishing NSI
parameters, i.e., $\Delta P_{e\mu}(\delta) = P^{\rm
SD}_{e\mu}(\delta) -P^{\rm SD}_{e\mu}(0)$ with $\delta=\pi/2$.}
\end{figure}
Finally, we show that the NSIs may lead to the enhancement of
CP-violating effects in neutrino oscillations \cite{Winter:2008eg}.
In Fig.~\ref{fig:CP}, the probability differences $\Delta
P_{e\mu}(\delta) \equiv P^{\rm NSI}_{e\mu}(\delta) - P^{\rm
NSI}_{e\mu}(0)$ induced by the leptonic CP violating phase $\delta$
are calculated for $\delta = \pi/2$. To signify the CP-violating
effects due to $\delta$, we set all the NSI phase parameters to
zero, i.e., $\phi_{\alpha \beta} = 0$. As mentioned before, the
CP-violating effects in the standard case come from the interference
between two different oscillation frequencies, and thus these
effects are relatively small in both cases of NH and IH. However,
the probability differences are enhanced greatly when the NSI
parameters are switched on, in particular in the NH case.

The left and right plots in the first row correspond to $\Delta
P_{e\mu}(\delta) = P^{\rm SD}_{e\mu}(\delta) - P^{\rm SD}_{e\mu}(0)$
in the NH and IH cases, respectively, since the NSI parameters are
taken to be zero. One can observe that the impact of $\delta$ is
insignificant. See Ref.~\cite{TZZ} for more discussions about the
CP-violating effects in the standard three-neutrino oscillations in
matter. Even if the NSI parameters are nonzero, the CP-violating
effects are negligible in the IH case, as shown in the middle and
lower plots in the right column. This is because the transition
probability $P^{\rm NSI}_{e\mu}$ itself is suppressed rather than
enhanced by matter effects. However, as in the NH case, $|\Delta
P_{e\mu}(\delta)|$ for $\delta = \pi/2$ could be $12\%$ in the
mantle, while as large as $30\%$ in the core, as illustrated in the
middle and lower plots in the left column. At the same time, without
the NSI effects, it is vanishingly small in both mantle and core
regions. Therefore, the future long-baseline oscillation
experiments, where neutrino beams traverse the Earth mantle, will
have excellent sensitivities to NSI-enhanced CP-violating effects.

The remarkable difference between standard and non-standard
CP-violating effects becomes clear, if we look at the last line of
Eq.~(21). More explicitly, $\Delta P_{e\mu}(\delta)$ is proportional
to $\alpha s_{13}$ in the standard case, while to
$|\varepsilon_{e\mu} | s_{13}$ or $|\varepsilon_{e\tau}| s_{13}$ in
the non-standard case. For $\varepsilon_{\alpha \beta} = 0.1$ and
$\alpha \approx 0.03$, the standard CP violation is suppressed by
one order of magnitude.

\section{Event Rates at PINGU}

In previous sections, we have explored the general features of NSI
effects on neutrino oscillations in the Earth matter at the level of
oscillation probabilities, which should have important implications
for atmospheric neutrino experiments. In order to study the NSI
effects in a realistic experiment, we have to calculate the number
of neutrino events in small bins of energies $\Delta E$ and zenith
angles $\Delta \cos\theta_z$ for a terrestrial detector. To this
end, we take a multi-megaton scale ice Cherenkov detector for
example, such as the proposed PINGU detector at the South
Pole~\cite{Koskinen:2011zz}. The main idea of this proposal is to
make the Deep Core of the IceCube detector denser to lower the
energy threshold down to a few GeV, implying possible precision
measurements of atmospheric neutrinos. Such an experimental setup
would be rather interesting in view of its great physics potential
for the determination of neutrino mass hierarchy, and the
oscillation parameters~\cite{Akhmedov:2012ah, Agarwalla:2012uj,
Franco:2013in, Ribordy:2013xea}. We refer the readers to
Ref.~\cite{Koskinen:2011zz} for a detailed description of the PINGU
experiment.

Atmospheric neutrinos interact with the nucleons via charged-current
interactions in the detector and produce energetic charged leptons,
which are radiating Cherenkov photons when propagating in ice. The
Cherenkov photons will be captured by the dense strings of
photomultipliers. The charged muons $\mu^\pm$ leave clear and long
tracks in the detector, so the experimental resolution to the
direction of muon neutrinos is much better than that of electron
neutrinos. Now, we make a rough estimate of the distribution of
$\nu_\mu$-like events collected by the PINGU detector for one year
running. Explicitly, the number of $\nu_\mu$-like events in the
$i$-th zenith-angle bin and $j$-th energy bin is given by
\begin{eqnarray}
N_{ij} &=& 2\pi N_{\rm A} T
\int_{\cos\theta^i_z}^{\cos\theta^{i+1}_z} {\rm d}\cos\theta_z
\int_{E_j}^{E_{j+1}} {\rm d}E \left[\left(\frac{{\rm
d}\Phi_{\nu_\mu}}{{\rm d}\cos\theta_z {\rm d}E} P^{\rm NSI}_{\mu\mu}
+ \frac{{\rm d}\Phi_{\nu_e}}{{\rm d}\cos\theta {\rm d}E}
P^{\rm NSI}_{e\mu} \right) \sigma^{\rm CC}_{\nu N}(E) \right. \nonumber \\
&& +\left. \left(\frac{{\rm d}\Phi_{\bar{\nu}_\mu}}{{\rm
d}\cos\theta {\rm d}E} \bar{P}^{\rm NSI}_{\mu\mu} + \frac{{\rm
d}\Phi_{\bar{\nu}_e}}{{\rm d}\cos\theta {\rm d}E} \bar{P}^{\rm
NSI}_{e\mu} \right) \sigma^{\rm CC}_{\bar{\nu}N}(E) \right] ~ \rho
V_{\rm eff}(E) \; , \label{eq:N}
\end{eqnarray}
where $\Phi_{\nu_\alpha}$ ($\Phi_{\bar{\nu}_\alpha}$) denotes the
neutrino $\nu_\alpha$ (antineutrino $\bar{\nu}_\alpha$) fluxes,
$N_{\rm A}$ is the Avogadro's number, $P^{\rm NSI}_{\alpha \beta}$
($\bar{P}^{\rm NSI}_{\alpha\beta})$ stands for the neutrino
(antineutrino) oscillation probabilities with NSI effects, and the
effective volume of PINGU with 20 strings is parametrized
as~\cite{Akhmedov:2012ah}
\begin{eqnarray}
\rho V_{\rm eff} (E)= 14.6~{\rm Mt} \times \left[\log_{10}
\left(\frac{E}{\rm GeV}\right)\right]^{1.8} \; .
\end{eqnarray}
Furthermore, we adopt the following simple parametrization of the
deep inelastic $\nu$-$N$ and $\bar{\nu}$-$N$ scattering cross
sections~\cite{Akhmedov:2012ah}
\begin{eqnarray}
\sigma^{\rm CC}_{\nu N}(E) &=& 7.30 \times 10^{-39}~{\rm cm}^2
\left(\frac{E}{\rm GeV}\right) \;, \nonumber
\\
\sigma^{\rm CC}_{\bar{\nu}N}(E) &=& 3.77 \times 10^{-39}~{\rm cm}^2
\left(\frac{E}{\rm GeV}\right) \; .
\end{eqnarray}
The atmospheric electron and muon neutrino fluxes are taken from
Ref.~\cite{Honda:1995hz}, where the neutrino fluxes have been
calculated using a hybrid method for $1~{\rm GeV} < E < 10^4~{\rm
GeV}$ and tabulated in small bins of neutrino energy and zenith
angle.

\begin{figure}[!t]\vspace{-0.3cm}
\includegraphics[width=.48\textwidth]{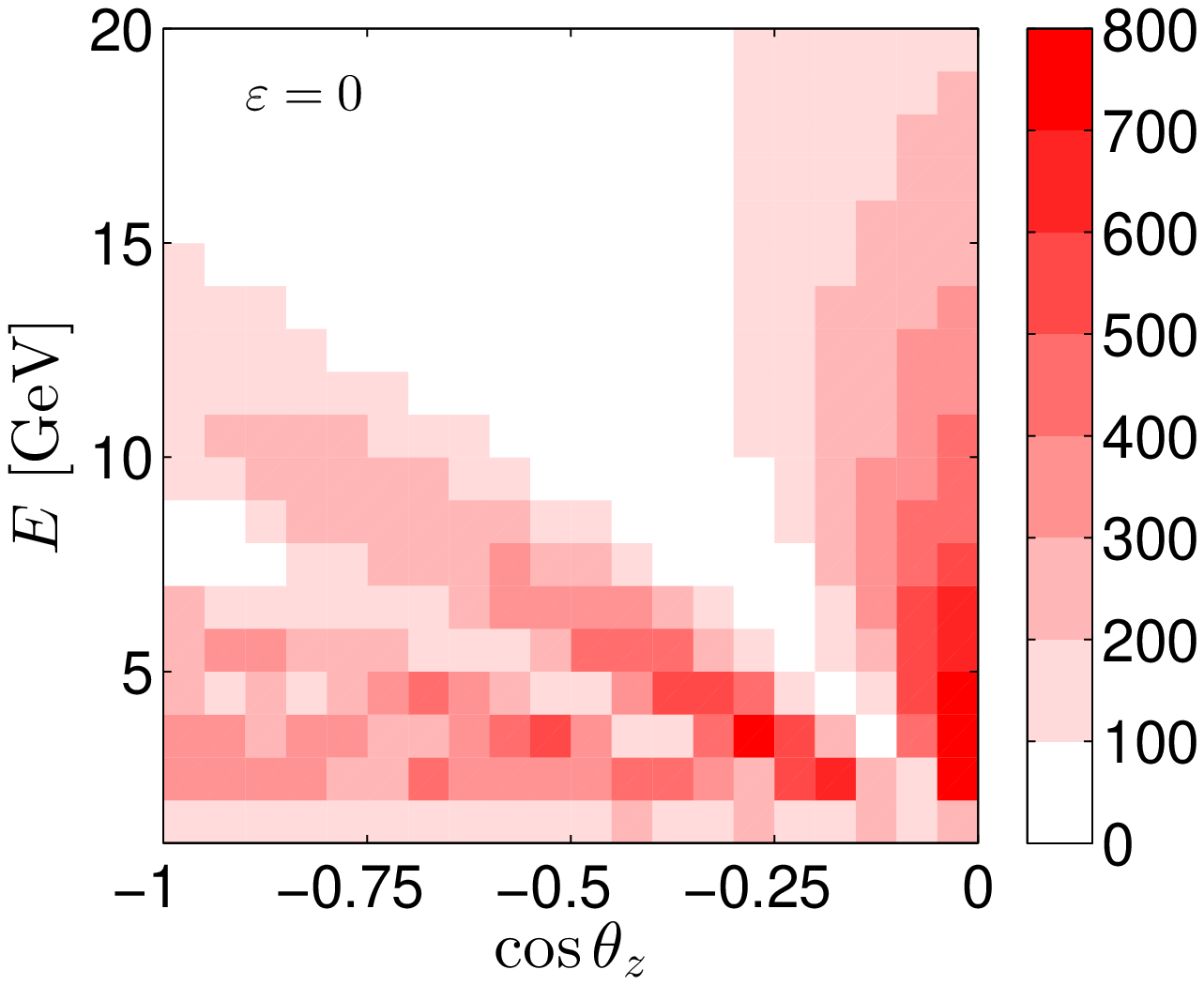}
\includegraphics[width=.48\textwidth]{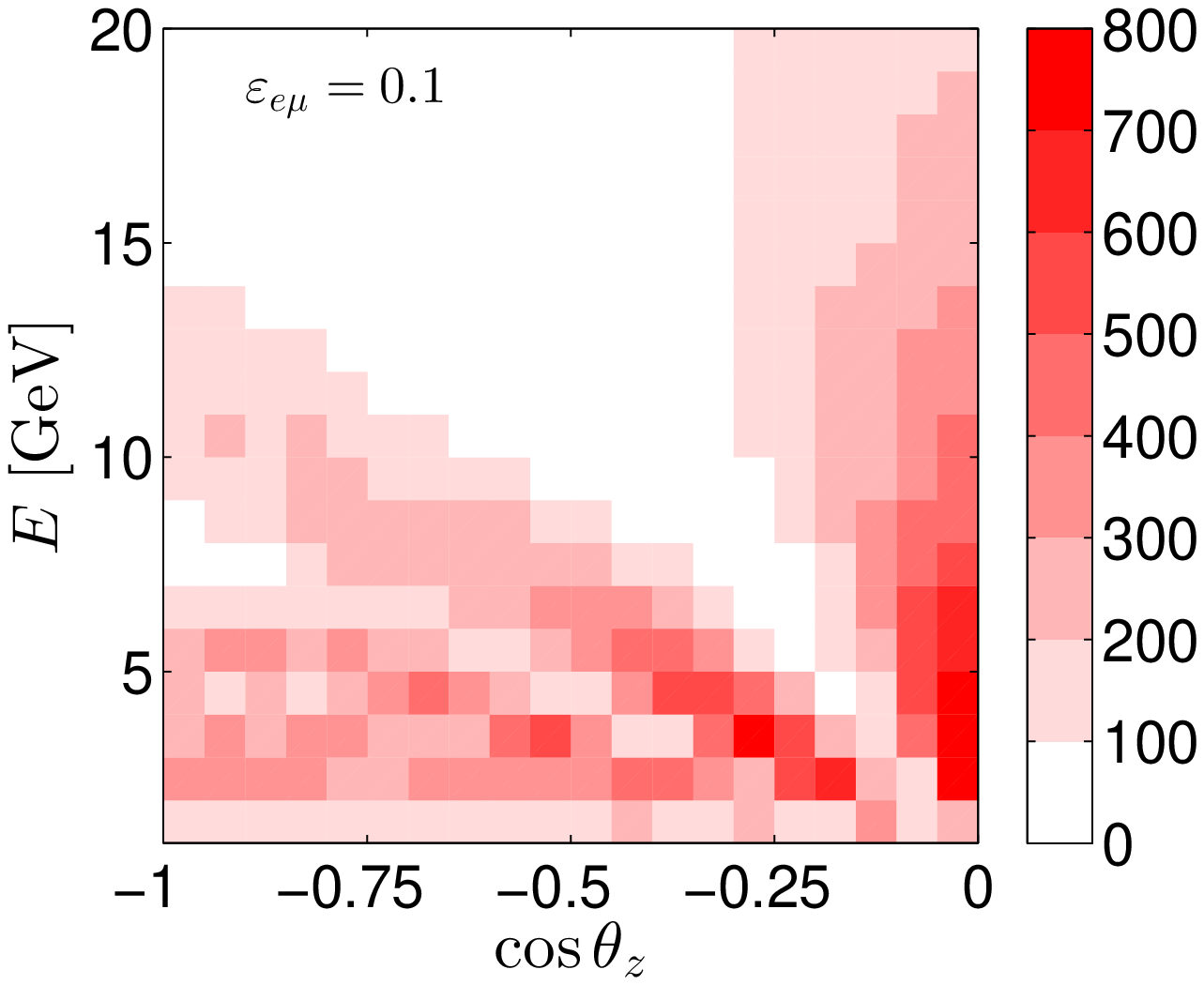}
\includegraphics[width=.48\textwidth]{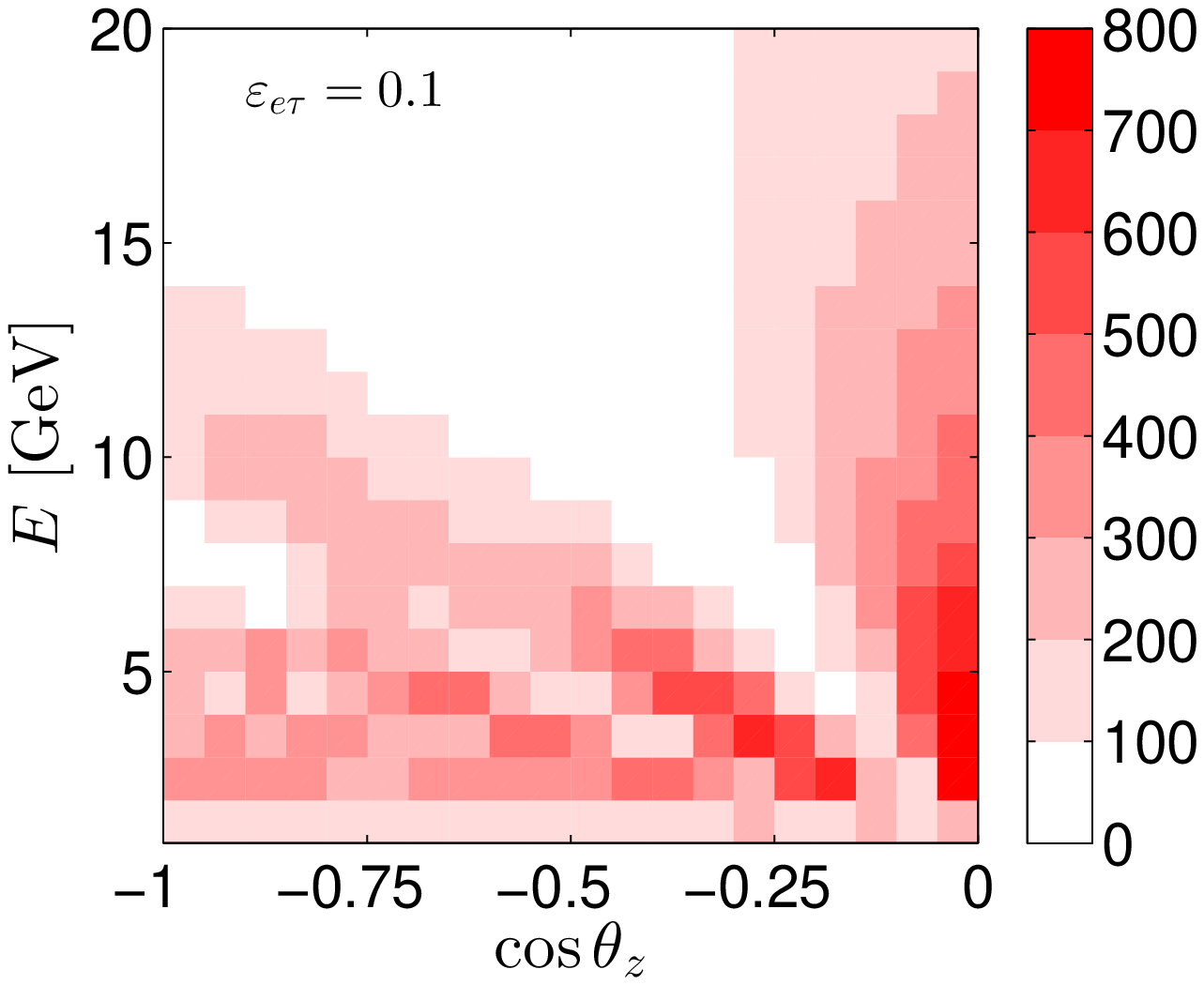}
\includegraphics[width=.48\textwidth]{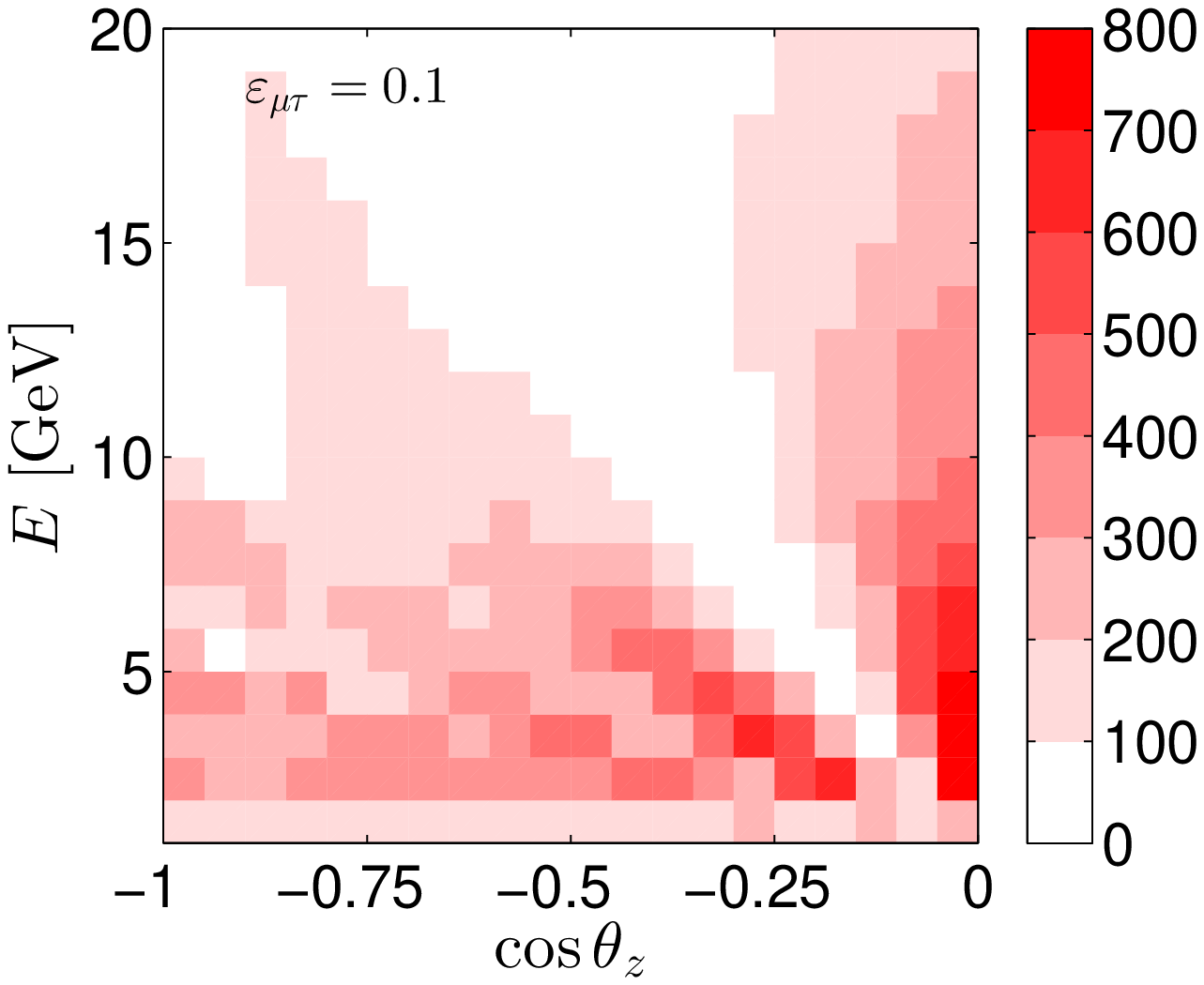}
\caption{\label{fig:N} The number of $\nu_\mu$-like events in the
$E-\cos\theta_z$ plane, collected by the PINGU detector for one
year, where the normal neutrino mass hierarchy is assumed. The upper
left plot corresponds to the case of standard neutrino oscillations,
i.e., all the NSI parameters are vanishing.}
\end{figure}

The event distribution in the case of standard neutrino oscillations
is shown in the upper left panel of Fig.~\ref{fig:N}, where the bin
widths $\Delta E=1~{\rm GeV}$ and $\Delta \cos\theta_z=0.05$ are
used and NH is assumed. Note that the highest number of events can
reach $\sim 800$, which is in agreement with the estimation in
Ref.~\cite{Akhmedov:2012ah}. It is evident that the event
distribution is mainly determined by the $\nu_\mu \to \nu_\mu$
survival probability, which is given in the left panel of
Fig.~\ref{fig:P2}. In the presence of NSIs, the event distribution
is distorted, as shown in the other plots of Fig.~\ref{fig:N}. In
particular, in the case of $\varepsilon_{\mu\tau} \neq 0$, the
number of events at high energy bins is increased remarkably. This
is because the $\nu_\mu \to \nu_\mu$ channel in Eq.~\eqref{eq:N}
dominates the contributions to the observed $\nu_\mu$-like events,
and the NSI parameter $\varepsilon_{\mu\tau}$ significantly modifies
the survival probability in this region. The latter has already been
pointed out in Sec. III C and illustrated in Fig.~\ref{fig:P2}.
Therefore, the PINGU detector has a better sensitivity to
$\varepsilon_{\mu\tau}$ than $\varepsilon_{e\mu}$ and
$\varepsilon_{e\tau}$. Note also that the total number of events is
approximately unchanged.

\begin{figure}[!h]\vspace{-0.3cm}
\includegraphics[width=.48\textwidth]{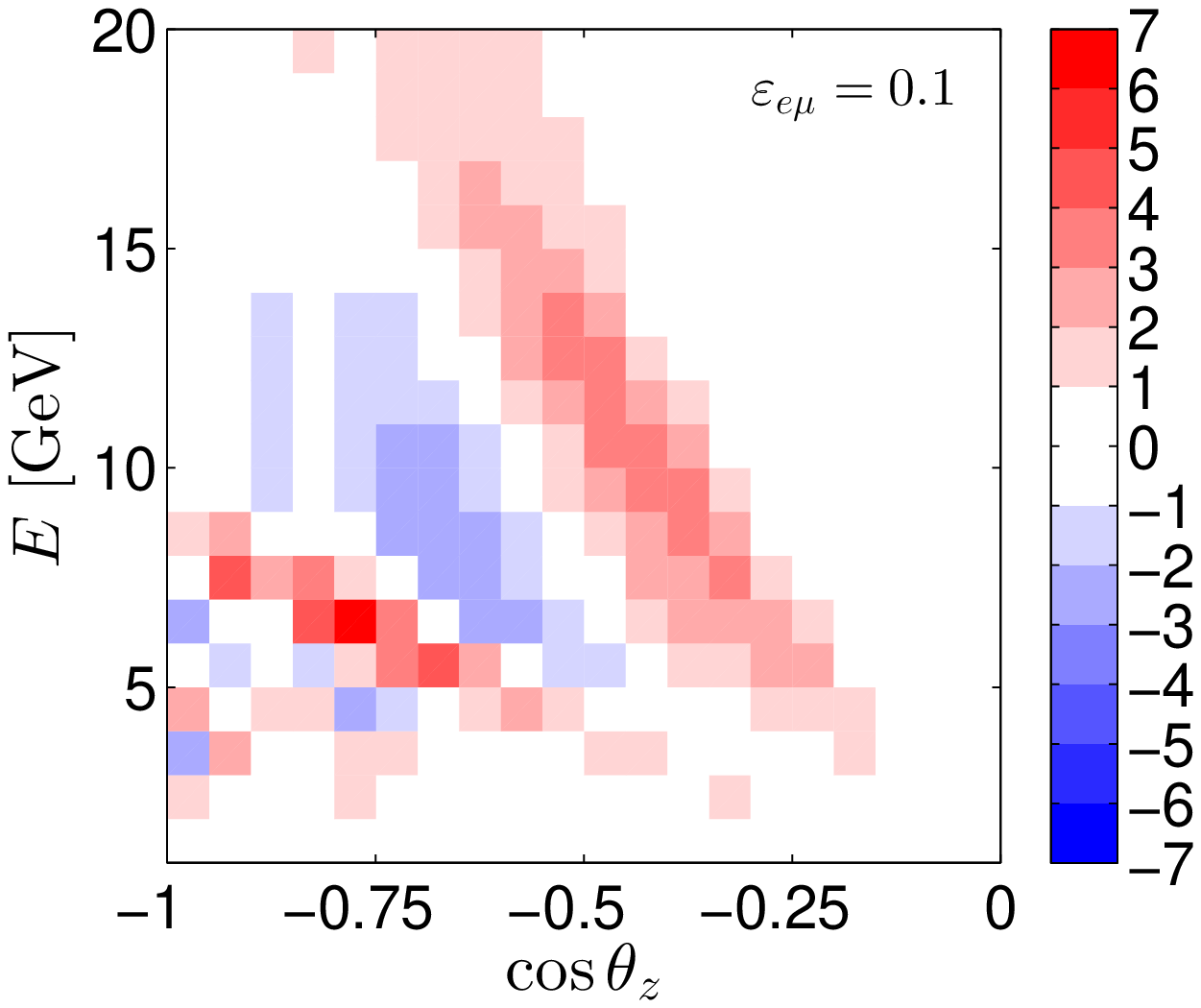}
\includegraphics[width=.48\textwidth]{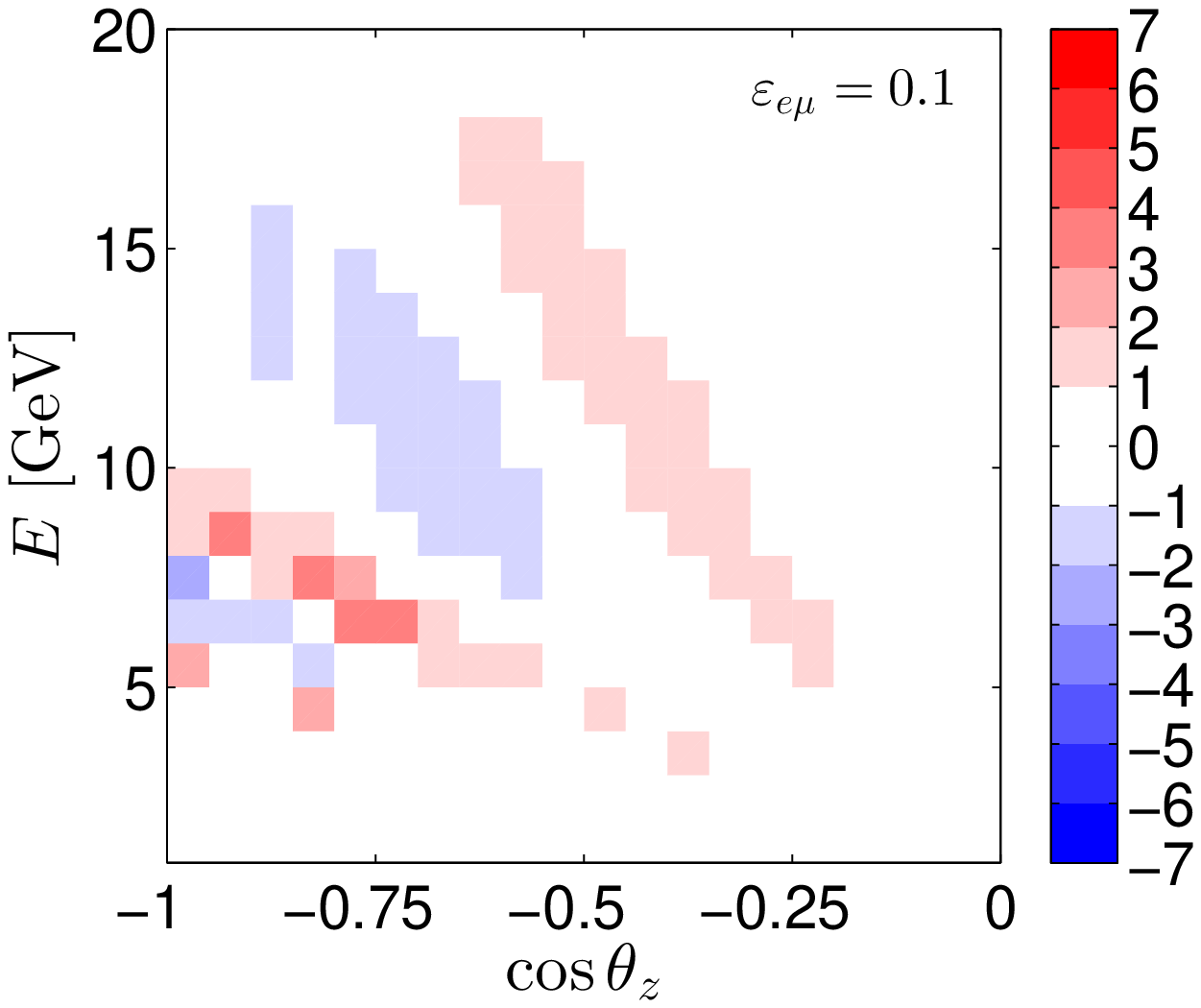}
\includegraphics[width=.48\textwidth]{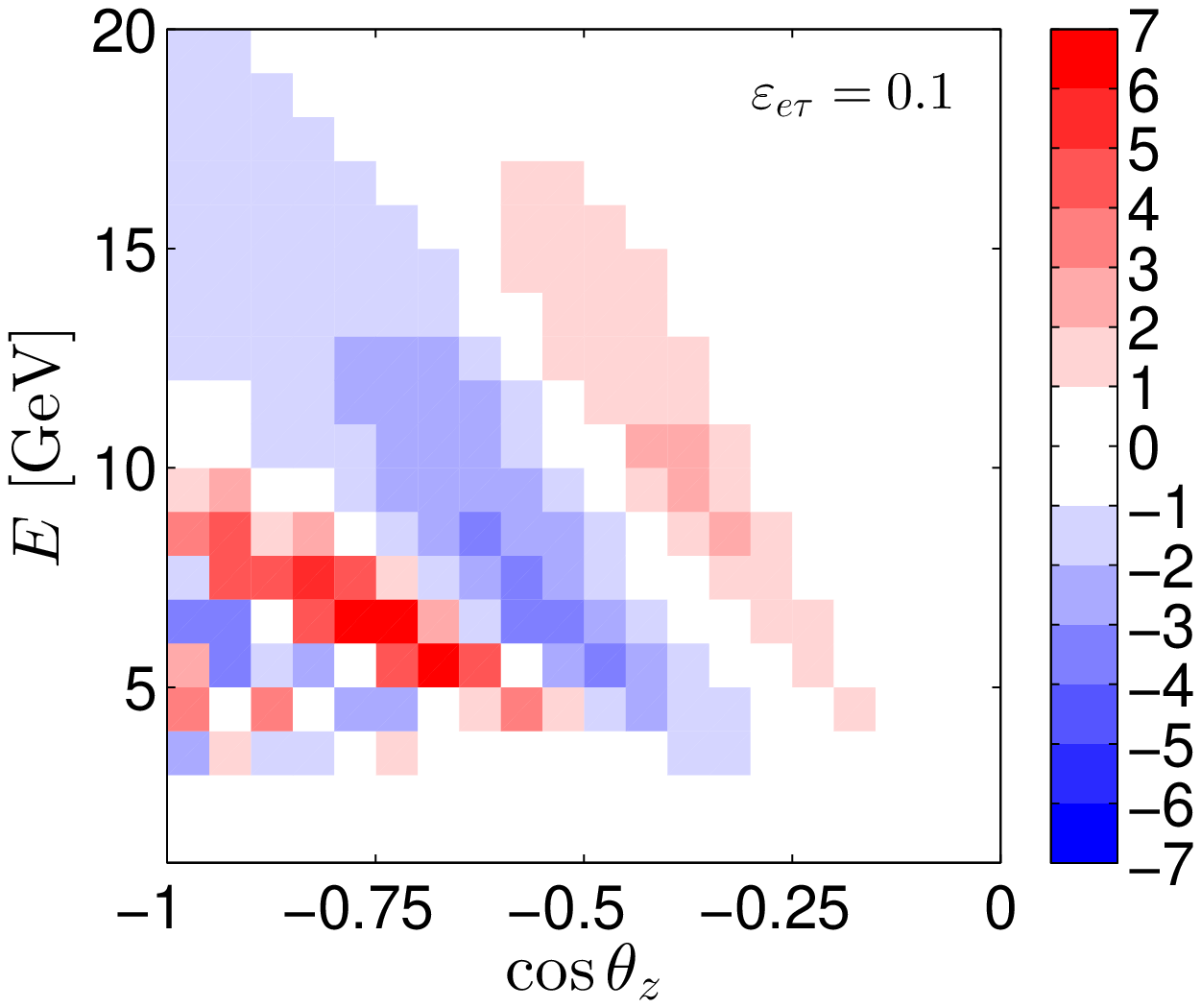}
\includegraphics[width=.48\textwidth]{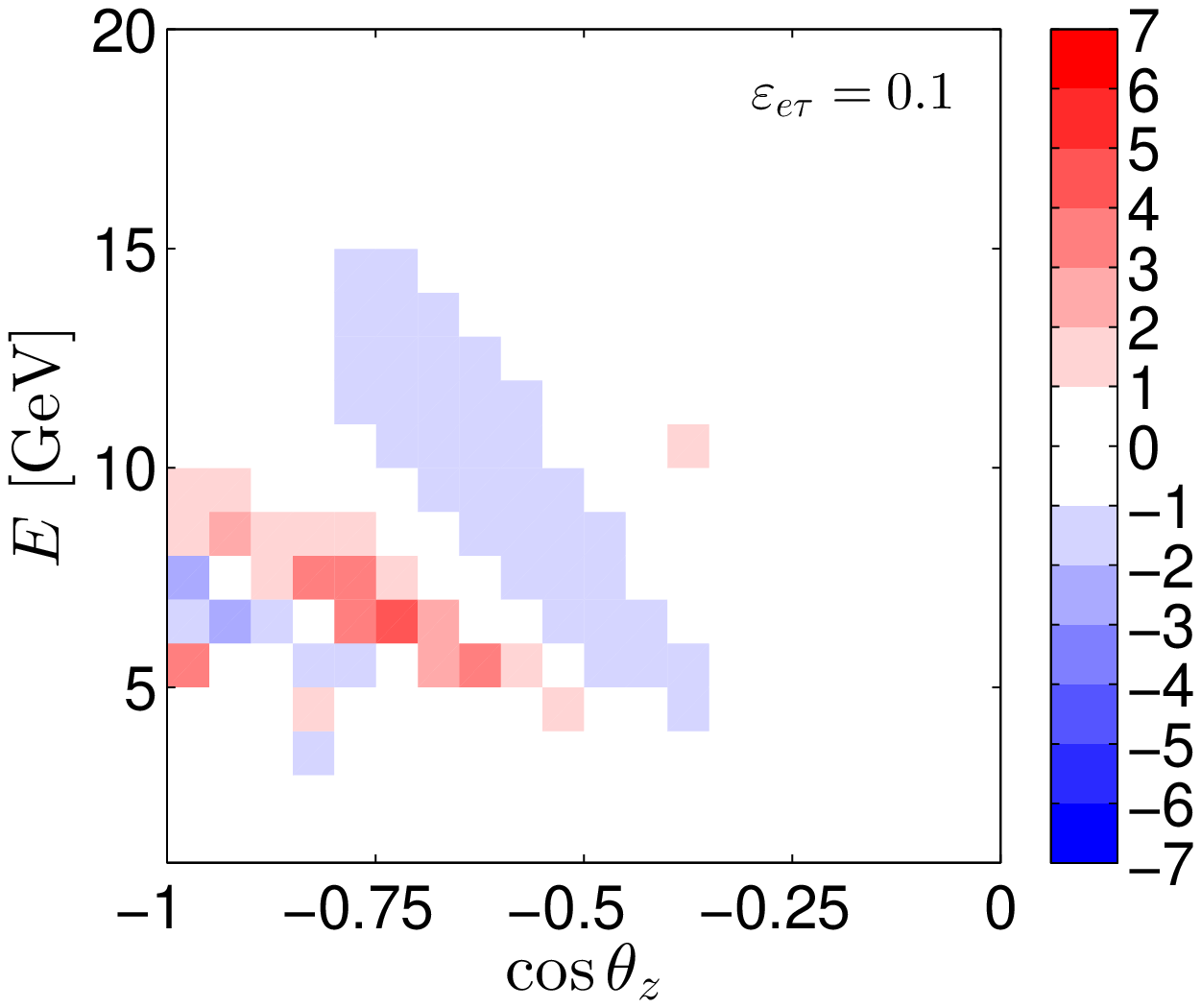}
\includegraphics[width=.48\textwidth]{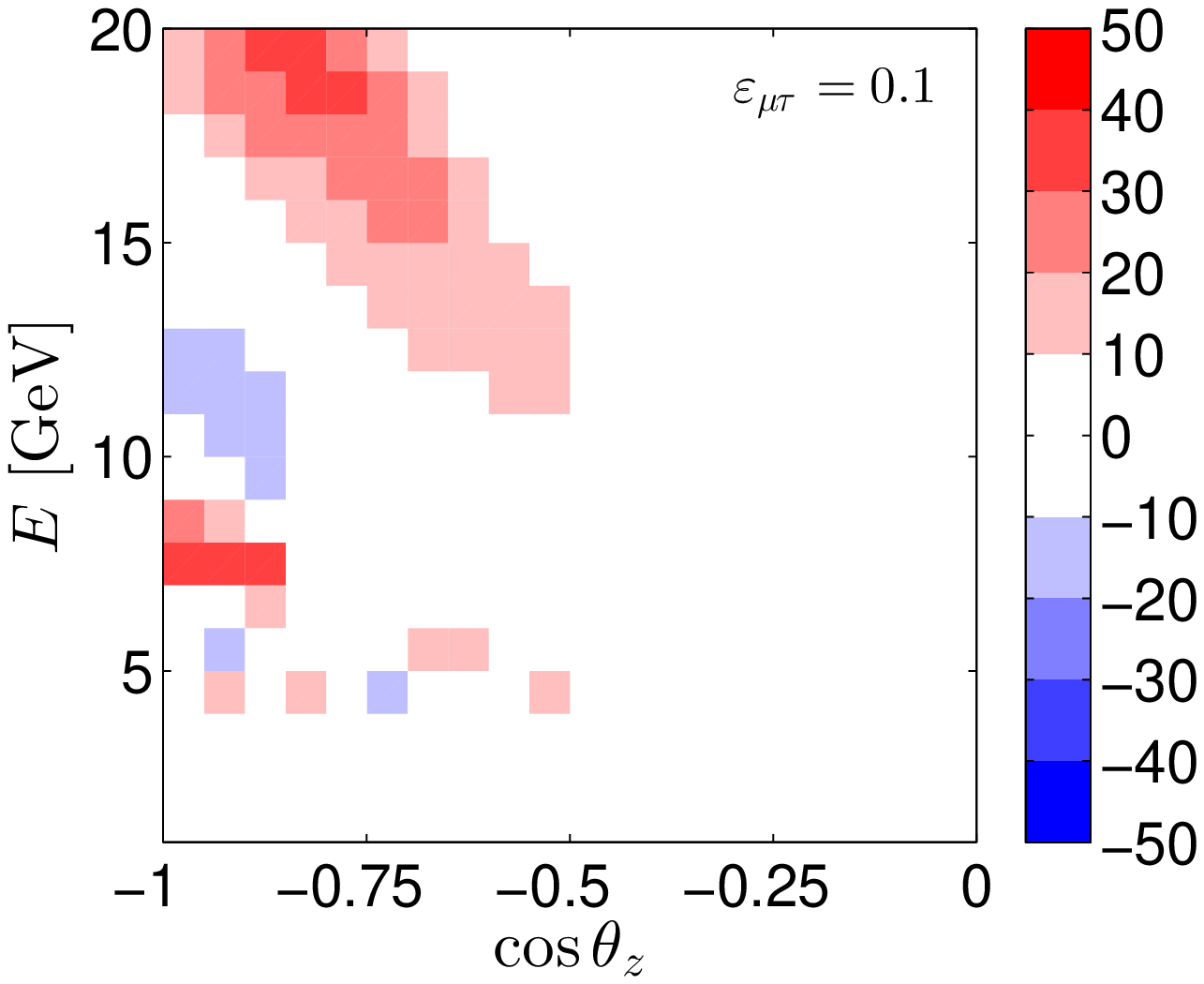}
\includegraphics[width=.48\textwidth]{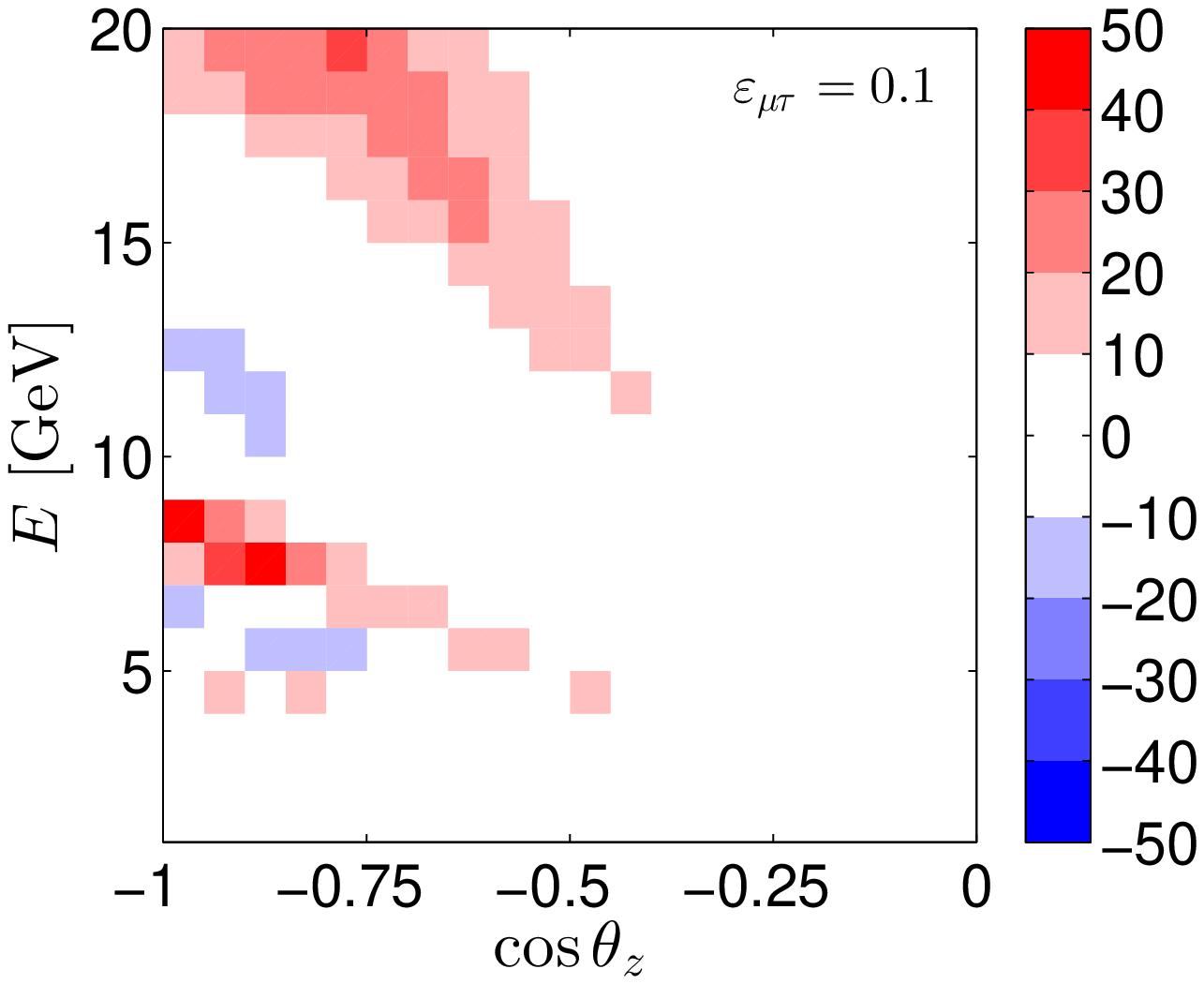}
\caption{\label{fig:S} The SD--NSI asymmetry ${\cal A} \equiv
(N^{\rm SD}_\mu - N^{\rm NSI}_\mu)/\sqrt{N^{\rm SD}_\mu}$ of
$\nu_\mu$-like events at the PINGU detector for one year, in the
cases of normal neutrino mass hierarchy (left column) and inverted
neutrino mass hierarchy (right column).}
\end{figure}

To quantify the significance of NSI effects, we define an SD--NSI
asymmetry as the difference between the number of $\nu_\mu$-like
events in the standard case (i.e., $\varepsilon_{\alpha\beta} = 0$)
and that in the NSI case
\begin{eqnarray}
{\cal A}=\frac{N^{\rm SD}_\mu-N^{\rm NSI}_\mu}{\sqrt{N^{\rm
SD}_\mu}}
\end{eqnarray}
for each bin in the $E$ -- $\cos\theta_z$ plane. In
Fig.~\ref{fig:S}, we illustrate the distribution of the asymmetry
$\cal A$ for specific values of NSI parameters. In the cases of
$\varepsilon_{e\mu} \neq 0$ and $\varepsilon_{e\tau} \neq 0$, the
large asymmetry appears in the low energy regions $E \sim
[5,10]~{\rm GeV}$, and is separated by the lines of zero asymmetry.
For $\varepsilon_{e\mu} = 0.1$ or $\varepsilon_{e\tau} = 0.1$, the
maximal asymmetry can be as large as ${\cal A} \approx 7$, which is
comparable to the estimated sensitivity to the neutrino mass
hierarchy~\cite{Akhmedov:2012ah}. As expected, in the case of a
non-vanishing $\varepsilon_{\mu\tau}$, the SD--NSI asymmetry is much
more stunning (e.g., ${\cal A} \approx 50$), indicating a great
discovery potential on this NSI parameter. We have also made a
comparison between NH and IH in Fig.~\ref{fig:S}, where one can
observe that the asymmetry in some bins is reduced in the case of
$\varepsilon_{e\mu} \neq 0$ and $\varepsilon_{e\tau} \neq 0$ for IH,
while there is no significant difference in the case of
$\varepsilon_{\mu\tau} \neq 0$. Therefore, the experimental
sensitivity to the NSI parameters, in particular
$\varepsilon_{\mu\tau}$, at the PINGU detector, is almost
independent of neutrino mass hierarchy.

Note that the real sensitivity to NSI effects will actually be much
lower because of systematics, the smearing effects in the
reconstruction of neutrino energies and zenith angles, the
uncertainties in the matter density profile and the other neutrino
oscillation parameters \cite{Akhmedov:2012ah}. In addition, the
$\nu_\tau$ interactions lead to an important background to the
$\nu_\mu$ events. Nevertheless, a detailed simulation on the probe
of NSIs at PINGU is meaningful and will be elaborated elsewhere.

\section{Summary}

Now we are entering a new era of precision measurements of neutrino
parameters, including three leptonic mixing angles and two neutrino
mass-squared differences. The ongoing and forthcoming neutrino
experiments are expected to pin down neutrino mass hierarchy and to
discover CP violation in the lepton sector. As the precisions of
oscillation experiments gradually improve and more data are
accumulated, it is promising to discover or constrain the new
physics effects beyond the standard paradigm of neutrino
oscillations.

One of the most widely studied scenarios is the non-standard
neutrino interactions. In this paper, we have considered the NSI
effects on neutrino oscillations in the Earth matter by using
neutrino oscillograms. First, we derive the mapping formulas between
the lepton mixing matrix in vacuum and that in matter beyond the
leading order approximation. In particular, the NSI effects on the
effective mixing angle $\tilde{\theta}_{13}$ have been discussed in
some detail. Then, the NSI effects on the neutrino oscillograms of
the Earth are investigated. The most significant difference between
the standard and non-standard oscillograms appears in the $\nu_\mu
\to \nu_\mu$ survival probability and in the case of
$\varepsilon_{\mu\tau} \neq 0$. In addition, the CP-violating
effects in neutrino oscillations can be enhanced by the NSI effects,
even if the CP-violating phases in the NSI parameters are switched
off. Finally, the NSI effects in the PINGU experiment are explored.
We calculate the event rate of atmospheric muon neutrinos at PINGU,
and demonstrate that the future huge atmospheric neutrino
experiments should have very good sensitivities to the NSI
parameters, in particular $\varepsilon_{\mu\tau}$. However, a more
sophisticated simulation of the NSI effects at PINGU, including the
systematics and other uncertainties, is needed to make a final
conclusion.

In addition to atmospheric neutrino experiments, such as PINGU, the
ongoing and upcoming long-baseline neutrino oscillation experiments,
which are intended for the determination of neutrino mass hierarchy
and leptonic CP violation, are also sensitive to the NSI effects. In
general, the future neutrino oscillation data will soon lead us
either to the discovery of new effects beyond the standard
oscillation scenario, or to more restrictive constraints on new
physics parameters.

\acknowledgements

One of the authors (H.Z.) is indebted to Evgeny Akhmedov for useful
discussions. This work was supported by the Swedish Research Council
(Vetenskapsr{\aa}det), contract no. 621-2011-3985 (T.O.), the Max
Planck Society through the Strategic Innovation Fund in the project
MANITOP (H.Z.), and the G\"{o}ran Gustafsson Foundation (S.Z.).

\bibliography{bibnsi}

\begin{thebibliography}{36}
\expandafter\ifx\csname natexlab\endcsname\relax\def\natexlab#1{#1}\fi
\expandafter\ifx\csname bibnamefont\endcsname\relax
  \def\bibnamefont#1{#1}\fi
\expandafter\ifx\csname bibfnamefont\endcsname\relax
  \def\bibfnamefont#1{#1}\fi
\expandafter\ifx\csname citenamefont\endcsname\relax
  \def\citenamefont#1{#1}\fi
\expandafter\ifx\csname url\endcsname\relax
  \def\url#1{\texttt{#1}}\fi
\expandafter\ifx\csname urlprefix\endcsname\relax\def\urlprefix{URL }\fi
\providecommand{\bibinfo}[2]{#2}
\providecommand{\eprint}[2][]{\url{#2}}

\bibitem[{\citenamefont{Beringer et~al.}(2012)}]{Beringer:1900zz}
\bibinfo{author}{\bibfnamefont{J.}~\bibnamefont{Beringer}} \bibnamefont{et~al.}
  (\bibinfo{collaboration}{Particle Data Group}), \bibinfo{journal}{Phys.~Rev.}
  \textbf{\bibinfo{volume}{D86}}, \bibinfo{pages}{010001}
  (\bibinfo{year}{2012}).

\bibitem[{\citenamefont{Ohlsson}(2013)}]{Ohlsson:2012kf}
\bibinfo{author}{\bibfnamefont{T.}~\bibnamefont{Ohlsson}},
  \bibinfo{journal}{Rep.~Prog.~Phys.} \textbf{\bibinfo{volume}{76}},
  \bibinfo{pages}{044201} (\bibinfo{year}{2013}), \eprint{1209.2710}.

\bibitem[{\citenamefont{Blennow et~al.}(2008)\citenamefont{Blennow, Ohlsson,
  and Skrotzki}}]{Blennow:2007pu}
\bibinfo{author}{\bibfnamefont{M.}~\bibnamefont{Blennow}},
  \bibinfo{author}{\bibfnamefont{T.}~\bibnamefont{Ohlsson}}, \bibnamefont{and}
  \bibinfo{author}{\bibfnamefont{J.}~\bibnamefont{Skrotzki}},
  \bibinfo{journal}{Phys.~Lett.} \textbf{\bibinfo{volume}{B660}},
  \bibinfo{pages}{522} (\bibinfo{year}{2008}), \eprint{hep-ph/0702059}.

\bibitem[{\citenamefont{Blennow and Ohlsson}(2008)}]{Blennow:2008eb}
\bibinfo{author}{\bibfnamefont{M.}~\bibnamefont{Blennow}} \bibnamefont{and}
  \bibinfo{author}{\bibfnamefont{T.}~\bibnamefont{Ohlsson}},
  \bibinfo{journal}{Phys.~Rev.} \textbf{\bibinfo{volume}{D78}},
  \bibinfo{pages}{093002} (\bibinfo{year}{2008}), \eprint{0805.2301}.

\bibitem[{\citenamefont{Fornengo et~al.}(2001)\citenamefont{Fornengo, Maltoni,
  Tomas, and Valle}}]{Fornengo:2001pm}
\bibinfo{author}{\bibfnamefont{N.}~\bibnamefont{Fornengo}},
  \bibinfo{author}{\bibfnamefont{M.}~\bibnamefont{Maltoni}},
  \bibinfo{author}{\bibfnamefont{R.}~\bibnamefont{Tomas}}, \bibnamefont{and}
  \bibinfo{author}{\bibfnamefont{J.}~\bibnamefont{Valle}},
  \bibinfo{journal}{Phys.~Rev.} \textbf{\bibinfo{volume}{D65}},
  \bibinfo{pages}{013010} (\bibinfo{year}{2001}), \eprint{hep-ph/0108043}.

\bibitem[{\citenamefont{Huber and Valle}(2001)}]{Huber:2001zw}
\bibinfo{author}{\bibfnamefont{P.}~\bibnamefont{Huber}} \bibnamefont{and}
  \bibinfo{author}{\bibfnamefont{J.}~\bibnamefont{Valle}},
  \bibinfo{journal}{Phys.~Lett.} \textbf{\bibinfo{volume}{B523}},
  \bibinfo{pages}{151} (\bibinfo{year}{2001}), \eprint{hep-ph/0108193}.

\bibitem[{\citenamefont{Mitsuka et~al.}(2011)}]{Mitsuka:2011ty}
\bibinfo{author}{\bibfnamefont{G.}~\bibnamefont{Mitsuka}} \bibnamefont{et~al.}
  (\bibinfo{collaboration}{Super-Kamiokande Collaboration}),
  \bibinfo{journal}{Phys.~Rev.} \textbf{\bibinfo{volume}{D84}},
  \bibinfo{pages}{113008} (\bibinfo{year}{2011}), \eprint{1109.1889}.

\bibitem[{\citenamefont{Kopp et~al.}(2008{\natexlab{a}})\citenamefont{Kopp,
  Lindner, Ota, and Sato}}]{Kopp:2007ne}
\bibinfo{author}{\bibfnamefont{J.}~\bibnamefont{Kopp}},
  \bibinfo{author}{\bibfnamefont{M.}~\bibnamefont{Lindner}},
  \bibinfo{author}{\bibfnamefont{T.}~\bibnamefont{Ota}}, \bibnamefont{and}
  \bibinfo{author}{\bibfnamefont{J.}~\bibnamefont{Sato}},
  \bibinfo{journal}{Phys.~Rev.} \textbf{\bibinfo{volume}{D77}},
  \bibinfo{pages}{013007} (\bibinfo{year}{2008}{\natexlab{a}}),
  \eprint{0708.0152}.

\bibitem[{\citenamefont{Ohlsson and Zhang}(2009)}]{Ohlsson:2008gx}
\bibinfo{author}{\bibfnamefont{T.}~\bibnamefont{Ohlsson}} \bibnamefont{and}
  \bibinfo{author}{\bibfnamefont{H.}~\bibnamefont{Zhang}},
  \bibinfo{journal}{Phys.~Lett.} \textbf{\bibinfo{volume}{B671}},
  \bibinfo{pages}{99} (\bibinfo{year}{2009}), \eprint{0809.4835}.

\bibitem[{\citenamefont{Leitner et~al.}(2011)\citenamefont{Leitner, Malinsky,
  Roskovec, and Zhang}}]{Leitner}
\bibinfo{author}{\bibfnamefont{R.}~\bibnamefont{Leitner}},
  \bibinfo{author}{\bibfnamefont{S.}~\bibnamefont{Malinsky}},
  \bibinfo{author}{\bibfnamefont{B.}~\bibnamefont{Roskovec}}, \bibnamefont{and}
  \bibinfo{author}{\bibfnamefont{H.}~\bibnamefont{Zhang}},
  \bibinfo{journal}{JHEP} \textbf{\bibinfo{volume}{1112}}, \bibinfo{pages}{001}
  (\bibinfo{year}{2011}), \eprint{1105.5580}.

\bibitem[{\citenamefont{Kopp et~al.}(2007)\citenamefont{Kopp, Lindner, and
  Ota}}]{Kopp:2007mi}
\bibinfo{author}{\bibfnamefont{J.}~\bibnamefont{Kopp}},
  \bibinfo{author}{\bibfnamefont{M.}~\bibnamefont{Lindner}}, \bibnamefont{and}
  \bibinfo{author}{\bibfnamefont{T.}~\bibnamefont{Ota}},
  \bibinfo{journal}{Phys.~Rev.} \textbf{\bibinfo{volume}{D76}},
  \bibinfo{pages}{013001} (\bibinfo{year}{2007}), \eprint{hep-ph/0702269}.

\bibitem[{\citenamefont{Ribeiro et~al.}(2007)\citenamefont{Ribeiro, Minakata,
  Nunokawa, Uchinami, and Zukanovich-Funchal}}]{Ribeiro:2007ud}
\bibinfo{author}{\bibfnamefont{N.}~\bibnamefont{Ribeiro}},
  \bibinfo{author}{\bibfnamefont{H.}~\bibnamefont{Minakata}},
  \bibinfo{author}{\bibfnamefont{H.}~\bibnamefont{Nunokawa}},
  \bibinfo{author}{\bibfnamefont{S.}~\bibnamefont{Uchinami}}, \bibnamefont{and}
  \bibinfo{author}{\bibfnamefont{R.}~\bibnamefont{Zukanovich-Funchal}},
  \bibinfo{journal}{JHEP} \textbf{\bibinfo{volume}{0712}}, \bibinfo{pages}{002}
  (\bibinfo{year}{2007}), \eprint{0709.1980}.

\bibitem[{\citenamefont{Kopp et~al.}(2008{\natexlab{b}})\citenamefont{Kopp,
  Ota, and Winter}}]{Kopp:2008ds}
\bibinfo{author}{\bibfnamefont{J.}~\bibnamefont{Kopp}},
  \bibinfo{author}{\bibfnamefont{T.}~\bibnamefont{Ota}}, \bibnamefont{and}
  \bibinfo{author}{\bibfnamefont{W.}~\bibnamefont{Winter}},
  \bibinfo{journal}{Phys.~Rev.} \textbf{\bibinfo{volume}{D78}},
  \bibinfo{pages}{053007} (\bibinfo{year}{2008}{\natexlab{b}}),
  \eprint{0804.2261}.

\bibitem[{\citenamefont{Meloni et~al.}(2010)\citenamefont{Meloni, Ohlsson,
  Winter, and Zhang}}]{Meloni:2009cg}
\bibinfo{author}{\bibfnamefont{D.}~\bibnamefont{Meloni}},
  \bibinfo{author}{\bibfnamefont{T.}~\bibnamefont{Ohlsson}},
  \bibinfo{author}{\bibfnamefont{W.}~\bibnamefont{Winter}}, \bibnamefont{and}
  \bibinfo{author}{\bibfnamefont{H.}~\bibnamefont{Zhang}},
  \bibinfo{journal}{JHEP} \textbf{\bibinfo{volume}{1004}}, \bibinfo{pages}{041}
  (\bibinfo{year}{2010}), \eprint{0912.2735}.

\bibitem[{\citenamefont{Coloma et~al.}(2011)\citenamefont{Coloma, Donini,
  Lopez-Pavon, and Minakata}}]{Coloma:2011rq}
\bibinfo{author}{\bibfnamefont{P.}~\bibnamefont{Coloma}},
  \bibinfo{author}{\bibfnamefont{A.}~\bibnamefont{Donini}},
  \bibinfo{author}{\bibfnamefont{J.}~\bibnamefont{Lopez-Pavon}},
  \bibnamefont{and} \bibinfo{author}{\bibfnamefont{H.}~\bibnamefont{Minakata}},
  \bibinfo{journal}{JHEP} \textbf{\bibinfo{volume}{1108}}, \bibinfo{pages}{036}
  (\bibinfo{year}{2011}), \eprint{1105.5936}.

\bibitem[{\citenamefont{Akhmedov et~al.}(2013)\citenamefont{Akhmedov, Razzaque,
  and Smirnov}}]{Akhmedov:2012ah}
\bibinfo{author}{\bibfnamefont{E.~K.} \bibnamefont{Akhmedov}},
  \bibinfo{author}{\bibfnamefont{S.}~\bibnamefont{Razzaque}}, \bibnamefont{and}
  \bibinfo{author}{\bibfnamefont{A.~Y.} \bibnamefont{Smirnov}},
  \bibinfo{journal}{JHEP} \textbf{\bibinfo{volume}{02}}, \bibinfo{pages}{082}
  (\bibinfo{year}{2013}), \eprint{1205.7071}.

\bibitem[{\citenamefont{Agarwalla et~al.}(2012)\citenamefont{Agarwalla, Li,
  Mena, and Palomares-Ruiz}}]{Agarwalla:2012uj}
\bibinfo{author}{\bibfnamefont{S.~K.} \bibnamefont{Agarwalla}},
  \bibinfo{author}{\bibfnamefont{T.}~\bibnamefont{Li}},
  \bibinfo{author}{\bibfnamefont{O.}~\bibnamefont{Mena}}, \bibnamefont{and}
  \bibinfo{author}{\bibfnamefont{S.}~\bibnamefont{Palomares-Ruiz}}
  (\bibinfo{year}{2012}), \eprint{1212.2238}.

\bibitem[{\citenamefont{Franco et~al.}(2013)\citenamefont{Franco, Jollet,
  Kouchner, Kulikovskiy, Meregaglia, Perasso, Pradier, Tonazzo, and
  Van~Elewyck}}]{Franco:2013in}
\bibinfo{author}{\bibfnamefont{D.}~\bibnamefont{Franco}},
  \bibinfo{author}{\bibfnamefont{C.}~\bibnamefont{Jollet}},
  \bibinfo{author}{\bibfnamefont{A.}~\bibnamefont{Kouchner}},
  \bibinfo{author}{\bibfnamefont{V.}~\bibnamefont{Kulikovskiy}},
  \bibinfo{author}{\bibfnamefont{A.}~\bibnamefont{Meregaglia}},
  \bibinfo{author}{\bibfnamefont{S.}~\bibnamefont{Perasso}},
  \bibinfo{author}{\bibfnamefont{T.}~\bibnamefont{Pradier}},
  \bibinfo{author}{\bibfnamefont{A.}~\bibnamefont{Tonazzo}}, \bibnamefont{and}
  \bibinfo{author}{\bibfnamefont{V.}~\bibnamefont{Van~Elewyck}},
  \bibinfo{journal}{JHEP} \textbf{\bibinfo{volume}{1304}}, \bibinfo{pages}{008}
  (\bibinfo{year}{2013}), \eprint{1301.4332}.

\bibitem[{\citenamefont{Ribordy and Smirnov}(2013)}]{Ribordy:2013xea}
\bibinfo{author}{\bibfnamefont{M.}~\bibnamefont{Ribordy}} \bibnamefont{and}
  \bibinfo{author}{\bibfnamefont{A.~Y.} \bibnamefont{Smirnov}}
  (\bibinfo{year}{2013}), \eprint{1303.0758}.

\bibitem[{\citenamefont{Akhmedov et~al.}(2001)\citenamefont{Akhmedov, Huber,
  Lindner, and Ohlsson}}]{Akhmedov:2001kd}
\bibinfo{author}{\bibfnamefont{E.~K.} \bibnamefont{Akhmedov}},
  \bibinfo{author}{\bibfnamefont{P.}~\bibnamefont{Huber}},
  \bibinfo{author}{\bibfnamefont{M.}~\bibnamefont{Lindner}}, \bibnamefont{and}
  \bibinfo{author}{\bibfnamefont{T.}~\bibnamefont{Ohlsson}},
  \bibinfo{journal}{Nucl.~Phys.} \textbf{\bibinfo{volume}{B608}},
  \bibinfo{pages}{394} (\bibinfo{year}{2001}), \eprint{hep-ph/0105029}.

\bibitem[{\citenamefont{Akhmedov et~al.}(2004)\citenamefont{Akhmedov,
  Johansson, Lindner, Ohlsson, and Schwetz}}]{Akhmedov:2004ny}
\bibinfo{author}{\bibfnamefont{E.~K.} \bibnamefont{Akhmedov}},
  \bibinfo{author}{\bibfnamefont{R.}~\bibnamefont{Johansson}},
  \bibinfo{author}{\bibfnamefont{M.}~\bibnamefont{Lindner}},
  \bibinfo{author}{\bibfnamefont{T.}~\bibnamefont{Ohlsson}}, \bibnamefont{and}
  \bibinfo{author}{\bibfnamefont{T.}~\bibnamefont{Schwetz}},
  \bibinfo{journal}{JHEP} \textbf{\bibinfo{volume}{0404}}, \bibinfo{pages}{078}
  (\bibinfo{year}{2004}), \eprint{hep-ph/0402175}.

\bibitem[{\citenamefont{Freund}(2001)}]{Freund:2001pn}
\bibinfo{author}{\bibfnamefont{M.}~\bibnamefont{Freund}},
  \bibinfo{journal}{Phys.~Rev.} \textbf{\bibinfo{volume}{D64}},
  \bibinfo{pages}{053003} (\bibinfo{year}{2001}), \eprint{hep-ph/0103300}.

\bibitem[{\citenamefont{Meloni et~al.}(2009)\citenamefont{Meloni, Ohlsson, and
  Zhang}}]{Meloni:2009ia}
\bibinfo{author}{\bibfnamefont{D.}~\bibnamefont{Meloni}},
  \bibinfo{author}{\bibfnamefont{T.}~\bibnamefont{Ohlsson}}, \bibnamefont{and}
  \bibinfo{author}{\bibfnamefont{H.}~\bibnamefont{Zhang}},
  \bibinfo{journal}{JHEP} \textbf{\bibinfo{volume}{0904}}, \bibinfo{pages}{033}
  (\bibinfo{year}{2009}), \eprint{0901.1784}.

\bibitem[{\citenamefont{Kikuchi et~al.}(2009)\citenamefont{Kikuchi, Minakata,
  and Uchinami}}]{Kikuchi:2008vq}
\bibinfo{author}{\bibfnamefont{T.}~\bibnamefont{Kikuchi}},
  \bibinfo{author}{\bibfnamefont{H.}~\bibnamefont{Minakata}}, \bibnamefont{and}
  \bibinfo{author}{\bibfnamefont{S.}~\bibnamefont{Uchinami}},
  \bibinfo{journal}{JHEP} \textbf{\bibinfo{volume}{0903}}, \bibinfo{pages}{114}
  (\bibinfo{year}{2009}), \eprint{0809.3312}.

\bibitem[{\citenamefont{Dziewonski and Anderson}(1981)}]{Dziewonski:1981xy}
\bibinfo{author}{\bibfnamefont{A.}~\bibnamefont{Dziewonski}} \bibnamefont{and}
  \bibinfo{author}{\bibfnamefont{D.}~\bibnamefont{Anderson}},
  \bibinfo{journal}{Phys.~Earth~Planet.~Interiors}
  \textbf{\bibinfo{volume}{25}}, \bibinfo{pages}{297} (\bibinfo{year}{1981}).

\bibitem[{\citenamefont{Davidson et~al.}(2003)\citenamefont{Davidson,
  Pena-Garay, Rius, and Santamaria}}]{Davidson:2003ha}
\bibinfo{author}{\bibfnamefont{S.}~\bibnamefont{Davidson}},
  \bibinfo{author}{\bibfnamefont{C.}~\bibnamefont{Pena-Garay}},
  \bibinfo{author}{\bibfnamefont{N.}~\bibnamefont{Rius}}, \bibnamefont{and}
  \bibinfo{author}{\bibfnamefont{A.}~\bibnamefont{Santamaria}},
  \bibinfo{journal}{JHEP} \textbf{\bibinfo{volume}{0303}}, \bibinfo{pages}{011}
  (\bibinfo{year}{2003}), \eprint{hep-ph/0302093}.

\bibitem[{\citenamefont{Biggio et~al.}(2009)\citenamefont{Biggio, Blennow, and
  Fernandez-Martinez}}]{Biggio:2009nt}
\bibinfo{author}{\bibfnamefont{C.}~\bibnamefont{Biggio}},
  \bibinfo{author}{\bibfnamefont{M.}~\bibnamefont{Blennow}}, \bibnamefont{and}
  \bibinfo{author}{\bibfnamefont{E.}~\bibnamefont{Fernandez-Martinez}},
  \bibinfo{journal}{JHEP} \textbf{\bibinfo{volume}{0908}}, \bibinfo{pages}{090}
  (\bibinfo{year}{2009}), \eprint{0907.0097}.

\bibitem[{\citenamefont{Adamson et~al.}(2013)}]{MINOS}
\bibinfo{author}{\bibfnamefont{P.}~\bibnamefont{Adamson}} \bibnamefont{et~al.}
  (\bibinfo{collaboration}{MINOS Collaboration}) (\bibinfo{year}{2013}),
  \eprint{1303.5314}.

\bibitem[{\citenamefont{Gonzalez-Garcia
  et~al.}(2012)\citenamefont{Gonzalez-Garcia, Maltoni, Salvado, and
  Schwetz}}]{GonzalezGarcia:2012sz}
\bibinfo{author}{\bibfnamefont{M.}~\bibnamefont{Gonzalez-Garcia}},
  \bibinfo{author}{\bibfnamefont{M.}~\bibnamefont{Maltoni}},
  \bibinfo{author}{\bibfnamefont{J.}~\bibnamefont{Salvado}}, \bibnamefont{and}
  \bibinfo{author}{\bibfnamefont{T.}~\bibnamefont{Schwetz}},
  \bibinfo{journal}{JHEP} \textbf{\bibinfo{volume}{1212}}, \bibinfo{pages}{123}
  (\bibinfo{year}{2012}), \eprint{1209.3023}.

\bibitem[{\citenamefont{Blennow et~al.}(2007)\citenamefont{Blennow, Ohlsson,
  and Winter}}]{Blennow:2005qj}
\bibinfo{author}{\bibfnamefont{M.}~\bibnamefont{Blennow}},
  \bibinfo{author}{\bibfnamefont{T.}~\bibnamefont{Ohlsson}}, \bibnamefont{and}
  \bibinfo{author}{\bibfnamefont{W.}~\bibnamefont{Winter}},
  \bibinfo{journal}{Eur.~Phys.~J.} \textbf{\bibinfo{volume}{C49}},
  \bibinfo{pages}{1023} (\bibinfo{year}{2007}), \eprint{hep-ph/0508175}.

\bibitem[{\citenamefont{Akhmedov et~al.}(2007)\citenamefont{Akhmedov, Maltoni,
  and Smirnov}}]{Akhmedov:2006hb}
\bibinfo{author}{\bibfnamefont{E.~K.} \bibnamefont{Akhmedov}},
  \bibinfo{author}{\bibfnamefont{M.}~\bibnamefont{Maltoni}}, \bibnamefont{and}
  \bibinfo{author}{\bibfnamefont{A.~Y.} \bibnamefont{Smirnov}},
  \bibinfo{journal}{JHEP} \textbf{\bibinfo{volume}{0705}}, \bibinfo{pages}{077}
  (\bibinfo{year}{2007}), \eprint{hep-ph/0612285}.

\bibitem[{\citenamefont{Akhmedov et~al.}(2008)\citenamefont{Akhmedov, Maltoni,
  and Smirnov}}]{Akhmedov:2008qt}
\bibinfo{author}{\bibfnamefont{E.~K.} \bibnamefont{Akhmedov}},
  \bibinfo{author}{\bibfnamefont{M.}~\bibnamefont{Maltoni}}, \bibnamefont{and}
  \bibinfo{author}{\bibfnamefont{A.~Y.} \bibnamefont{Smirnov}},
  \bibinfo{journal}{JHEP} \textbf{\bibinfo{volume}{0806}}, \bibinfo{pages}{072}
  (\bibinfo{year}{2008}), \eprint{0804.1466}.

\bibitem[{\citenamefont{Winter}(2009)}]{Winter:2008eg}
\bibinfo{author}{\bibfnamefont{W.}~\bibnamefont{Winter}},
  \bibinfo{journal}{Phys.~Lett.} \textbf{\bibinfo{volume}{B671}},
  \bibinfo{pages}{77} (\bibinfo{year}{2009}), \eprint{0808.3583}.

\bibitem[{\citenamefont{Ohlsson et~al.}(2013)\citenamefont{Ohlsson, Zhang, and
  Zhou}}]{TZZ}
\bibinfo{author}{\bibfnamefont{T.}~\bibnamefont{Ohlsson}},
  \bibinfo{author}{\bibfnamefont{H.}~\bibnamefont{Zhang}}, \bibnamefont{and}
  \bibinfo{author}{\bibfnamefont{S.}~\bibnamefont{Zhou}},
  \bibinfo{journal}{Phys.~Rev.} \textbf{\bibinfo{volume}{D87}},
  \bibinfo{pages}{053006} (\bibinfo{year}{2013}), \eprint{1301.4333}.

\bibitem[{\citenamefont{Koskinen}(2011)}]{Koskinen:2011zz}
\bibinfo{author}{\bibfnamefont{D.~J.} \bibnamefont{Koskinen}},
  \bibinfo{journal}{Mod.~Phys.~Lett.} \textbf{\bibinfo{volume}{A26}},
  \bibinfo{pages}{2899} (\bibinfo{year}{2011}).

\bibitem[{\citenamefont{Honda et~al.}(1995)\citenamefont{Honda, Kajita,
  Kasahara, and Midorikawa}}]{Honda:1995hz}
\bibinfo{author}{\bibfnamefont{M.}~\bibnamefont{Honda}},
  \bibinfo{author}{\bibfnamefont{T.}~\bibnamefont{Kajita}},
  \bibinfo{author}{\bibfnamefont{K.}~\bibnamefont{Kasahara}}, \bibnamefont{and}
  \bibinfo{author}{\bibfnamefont{S.}~\bibnamefont{Midorikawa}},
  \bibinfo{journal}{Phys.~Rev.} \textbf{\bibinfo{volume}{D52}},
  \bibinfo{pages}{4985} (\bibinfo{year}{1995}), \eprint{hep-ph/9503439}.

\end{thebibliography}

\end{document}